\documentclass[
11pt,
tightenlines,
longbibliography,
showpacs,
nofootinbib,
notitlepage,
superscriptaddress]{revtex4-2}

\usepackage{graphicx, multirow}
\usepackage{amsmath, amssymb, amsthm, physics}
\usepackage{algorithm, algorithmic}
\usepackage{tikz, hyperref}
\usepackage{booktabs} 
\usepackage{xtab} 
\usepackage{multirow} 
\usepackage{array}    
\usepackage{color}

\usepackage[capitalize]{cleveref}
\usepackage[a4paper, top=1in, bottom=1in, left=0.65in, right=0.65in]{geometry}

\usepackage{enumitem}
\setlist[itemize]{itemsep=0pt}

\AtBeginDocument{%
  \mathcode`l="8000
  \begingroup\lccode`\~=`\l
  \lowercase{\endgroup\edef~}{\noexpand\ell}%
}

\usepackage[font=scriptsize]{subcaption}
\usepackage[font=scriptsize, justification= raggedright]{caption}
\begin{document}

\title{Optimizing Multi-level Magic State Factories\\for Fault-Tolerant Quantum Architectures}

\author{Allyson~Silva}
\author{Artur~Scherer}
\author{Zak~Webb}
\author{Abdullah~Khalid}
\author{Bohdan~Kulchytskyy}
\author{Mia~Kramer}
\author{Kevin~Nguyen}
\author{Xiangzhou~Kong}
\author{Gebremedhin~A.~Dagnew}
\author{Yumeng~Wang}
\author{Huy~Anh~Nguyen}
\author{Einar~Gabbassov}
\author{Katiemarie~Olfert}
\affiliation{1QB Information Technologies (1QBit), Vancouver, BC, Canada}

\author{Pooya~Ronagh}
\thanks{{\vskip-10pt}{\hskip-8pt}Corresponding author: \href{mailto:pooya.ronagh@1qbit.com}{pooya.ronagh@1qbit.com}\\}
\affiliation{1QB Information Technologies (1QBit), Vancouver, BC, Canada}
\affiliation{Institute for Quantum Computing, University of Waterloo, Waterloo, ON, Canada}
\affiliation{Department of Physics \& Astronomy, University of Waterloo, Waterloo, ON, Canada}
\affiliation{Perimeter Institute for Theoretical Physics, Waterloo, ON, Canada}

\date{\today}

\begin{abstract}
We propose a novel technique for optimizing a modular fault-tolerant quantum
computing architecture, taking into account any desired space--time trade-offs
between the number of physical qubits and the fault-tolerant execution time of
a quantum algorithm. We consider a concept architecture comprising a dedicated
zone as a multi-level magic state factory and a core processor for efficient
logical operations, forming a supply chain network for production and consumption
of magic states. Using a heuristic algorithm, we solve the multi-objective
optimization problem of minimizing space and time subject to a user-defined
error budget for the success of the computation, taking the performance of
various fault-tolerant protocols into account. As an application, we show that
physical quantum resource estimation reduces to a simple model involving a
small number of key parameters, namely, the circuit volume, the error prefactors
($\mu$) and error suppression rates ($\Lambda$) of the fault-tolerant protocols,
the reaction time ($\gamma$), and an allowed slowdown factor ($\beta$).
\end{abstract}

\maketitle

\section{Introduction}

Fault-tolerant compilation of quantum algorithms is especially more complicated
than that of classical computer programs because the final physical circuit
depends on the specific noise characteristics of a quantum processing unit
(QPU). This information is required to decide the types and distances of error
correcting codes used, and to thereby estimate the fidelities of various
fault-tolerant protocols performed during the computation. Since utility-scale
computations may require $10^6$--$10^{12}$ non-Clifford operations performed
during several days or even months on complex and expensive fault-tolerant
quantum computing (FTQC) stacks, it is crucial to minimize the overhead of
fault tolerance while bounding the accumulated errors of logical operations
with high confidence.

In this paper, we provide a systematic approach to optimizing the design and
sizes of various zones of a modular FTQC architecture. We call this procedure
the \emph{assembly} of an FTQC program, which has been previously compiled from
a higher-level programming environment to a quantum assembly or
intermediate-representation (IR) language (such as QASM~\cite
{cross2022openqasm} and QIR~\cite{qir2024}). The key idea is to determine the
size of each FTQC architecture zone by choosing the number of FTQC protocol
units (e.g., 15:1 magic state distillation units, or the like) in that zone in
such a way that the supply and demand of quantum states between the
interconnected zones are balanced. This balance can be broken if a smaller total
number of physical qubits or a faster execution time is desired.

We first present a framework for constructing flexible varieties of FTQC
architectures in a modular fashion by allocating dedicated zones across a 2D
physical qubit layout for the execution of specific fault-tolerant protocols.
To this end, we view FTQC as the continual production and consumption of
various types of expensive quantum resources, such as the production of
lower-fidelity logical $\ket{T}$ or $\ket{CCZ}$ states in one zone and their
consumption in another zone for obtaining higher-fidelity magic states via a
distillation protocol~\citep{Gidney_2019_CCZ_distill}. We use rotated surface
codes as the typical choice of quantum error correction (QEC) codes, and rely
on lattice surgery for the execution of entangling gates and routing of quantum
information across a \textit{quantum bus}~\citep
{silva2024multi, litinski2018game}. We also introduce \emph{buffer registers}
between the zones to faciliate asynchronous operation of interconnected zones.
Such asynchronous operations are caused by differences in logical cycle times
across the architecture and the stochastic nature of magic state distillation.
The buffer registers ensure that the quantum computer operates in a long-term
steady state, and reduce the chance of execution-time halting. Newly produced
resources in one zone are stored and fault-tolerantly maintained via continued
stabilizer measurements and decoding (quantum memory) in the buffers until
their consumption in a subsequent zone. One or many terminal \textit
{core processors} consume the final highest-fidelity resources to perform the
logical gates of the quantum program. We refer the reader
to \cref{sec:architecture} and \cref{fig:core_area,fig:magic_state_factory} for
further details on the proposed architecture.

Our process shines particularly well when an MSF consists of heterogeous zones
with various code distances or distillation protocols. In contrast to previous
designs such as that of Ref.~\cite{beverland2022assessing}, which reuse the
same area in the MSF for sequential distillation, using dedicated zones for
distillation reduces the number of distillation units required in the MSF since
less time is required for distillation. Dedicated zones also allow failed
protocols to immediately and asynchronously restart, increasing the utilization
of resources (see \cref{app:comparison} for further comparative
discussions). Our assembly method can optimize the distances of each of
the levels of the MSF as shown in \cref{sec:solution}. Moreover, an intentional
oversupply or undersupply of resource states between the zones can be used to
tune various space--time trade-offs for FTQC. For example, a magic state
factory (MSF) may require multiple magic state distillation units to keep the
core processor fully stacked, especially when a reaction-time-limited
compilation is used to expedite the consumption rate of magic states \cite
{fowler2013time}. A smaller number of distillation units may still be able to
fault-tolerantly execute the quantum algorithm if some slowdown caused by
occasionally empty buffer registers can be tolerated. However, reducing the
(logical) size of the MSF increases the idle time in the core processor, which
increases the overall accumulated error rates, potentially requiring larger
code distances. Therefore it is not a priori obvious that such a strategy will
allow for longer runtime on a smaller quantum computer without taking the noise
profile of the hardware into account. Our study provides a detailed approach to
overcoming the challenges of analyzing such trade-offs.

We model the optimization of space and time as a bi-objective optimization
problem subject to a total error budget constraint discussed in
\cref{sec:exec_time} and \cref{app:error-model}. The problem is solved using
a simple and effective heuristic that determines the sizes of all levels of the
MSF and the code distances required by all logical qubits in the MSF and the
core processor, while meeting a total accumulated error budget. By solving this
optimization problem, we observe that the space and time costs of any FTQC can
be efficiently predicted from a small set of key attributes of (a) the quantum
program, and (b) the quantum computer executing it.
Firstly, the two relevant attributes
of the FTQC program are denoted by the Greek small letters $\alpha$ and
$\beta$, representing the average size of lattice surgeries in the core
processor relative to the number of logical qubits and a user-defined slowdown
factor within the buffers, respectively. Secondly, the noise
profile of the QPU is also important for the assembler. We observe that this
information can also be reduced to a small number of parameters
for a predictive model of the logical error rates of each FTQC protocol,
denoted by $\mu$ and $\Lambda$, respectively representing the logical error
prefactor and the logical error suppression rate of the protocol. Prior
literature have confined to using the quantum memory on a single surface code
patch, $\Lambda = \Lambda_{\text{mem}}$, to approximate the
performance of all lattice surgeries in the FTQC
program~\citep{acharya2024quantum}. However, a more comprehensive predictive
model must incorporate the exponential logical error suppression factors of all
fault-tolerant protocols in the assembly. Note that the
quantum memory error suppression rate $\Lambda_{\text{mem}}$ describes the
asymptotic multiple of improvement in the logical fidelity of the QEC code when
increasing its distance by 2. Therefore, in \Cref{app:error-model} we develop
more general logical error rate models for lattice surgeries with various shapes
and involving various numbers of surface code patches.

The paper is organized as follows. \Cref{sec:architecture} introduces our proposed architecture. \Cref{sec:exec_time} formalizes the bi-objective optimization problem to be solved when assembling the proposed architecture and the optimization framework we used to solve the problem. In \Cref{sec:experiments} we present several applications of our systematic assembly procedure by numerically studying FTQC execution space--time trade-offs with respect to specifications of the input quantum circuits, the quantum hardware noise profile, and the reaction time of classical decoders. Lastly, our concluding remarks follow in \Cref{sec:conclusion}.

\section{The Fault-Tolerant Architecture}
\label{sec:architecture}

We assume rotated surface codes as our QEC scheme of choice and assume that the program to be run is transpiled to multi-qubit Pauli rotations of the angles $\pi$, $\pi/2$, and $\pi/4$ for Clifford operations, and $\pi/8$ for non-Clifford operations~\citep{bravyi2016trading,litinski2019magic}. These Pauli rotations are in turn implemented using fault-tolerant lattice surgery~\citep{litinski2018game,beverland2022assessing,silva2024multi}. A core processor is designed to sequentially perform multi-qubit, long-range entanglements for sets of Pauli rotation measurements using auxiliary qubits to connect all the logical qubits required by a given logical operation. Clifford operations can be implemented via lattice surgery using the logical qubits involved with an auxiliary qubit initialized in the $\ket{0}$ state, while non-Clifford operations require entangling the logical qubits with a qubit in a magic state~\citep{litinski2018game}. The core (see \cref{sec:core_area}) is connected to an external multi-level MSF, as detailed in \cref{sec:magic_state_factory}, that supplies high-fidelity magic states required for the implementation of the non-Clifford logical operations.

For simplicity, we use an architecture with a single core processor and single MSF as our working example, although our analysis can be directly applied to more-sophisticated architectures with zones for creating other resource states such as the Toffoli~\citep{Eastin_2013_toffoli_distill} and doubly controlled $Z$ ($CCZ$)~\citep{Gidney_2019_CCZ_distill} magic state distillation units. Moreover, if a quantum algorithm uses a particular rotation angle frequently, a dedicated zone for specific rotation angles can be allocated. It may also be favourable to use distillation protocols that produce more magic states at the cost of a smaller increase in fidelity. Lastly, it might be useful to use dedicated zones for quantum subroutines such as QROM or arithmetic.

\subsection{The Core Processor} \label{sec:core_area}

The core processor is the central component of the architecture, where logical operations are performed on the logical qubits. It comprises two key elements: the memory fabric and a buffer.
The memory fabric is responsible for storing the quantum information required by the computation and performing its logical operations, while the buffer is used to simplify access to magic states so that the memory fabric does not have to directly interface with the MSF. There, the resources required by the $\pi/8$ rotations are prepared and held until they are consumed.

\begin{figure}[h]
    \centering
    \includegraphics[width=0.6\linewidth]{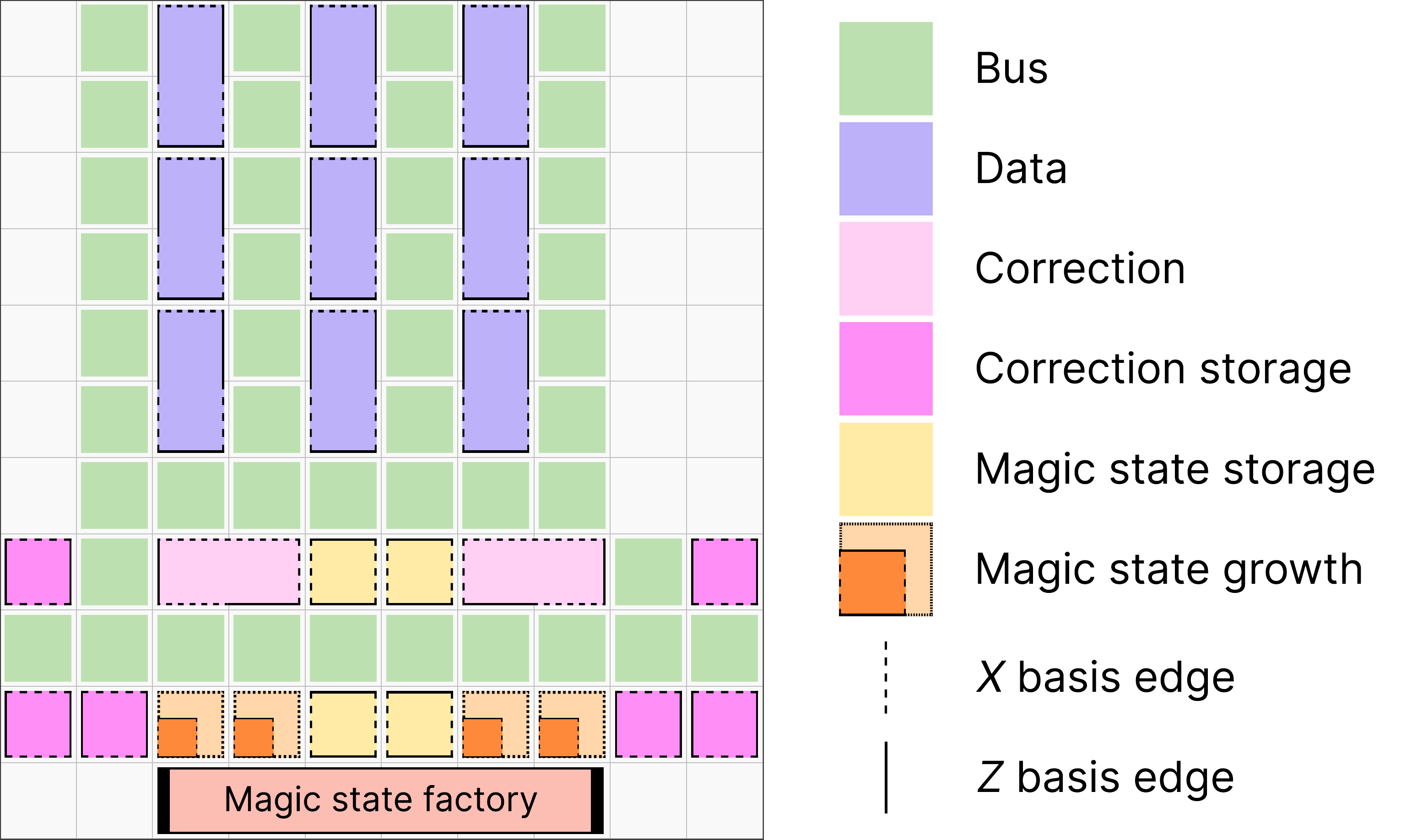}
    \caption{{\bf Example layout of the core processor.} The memory fabric is composed of $Q = 18$ logical qubits distributed in nine two-tile, two-qubit data patches surrounded by bus patches, while the buffer is composed of patches dedicated to storing magic states produced in the MSF. This storage space is designed to allow post-corrected $\pi/8$ rotations by connecting the magic state growth patches to a correction qubit patch using lattice surgery in parallel to the lattice surgery used to connect the data patches to the magic state consumed in a magic state storage patch. Magic state growth patches also allow magic state expansion if the produced magic states are of a different size than is required. The sides of the buffer are used to store correction qubits. The number of correction storage patches is determined by the time required to perform the classical processing in the post-corrected protocol.}
    \label{fig:core_area}
\end{figure}

An example layout of the core processor is shown in \cref{fig:core_area}\, for a circuit requiring 18 logical qubits to store its data. We follow the fast block layout of Ref.~\cite{litinski2018game} for the memory fabric, where logical qubits are arranged in two-tile, two-qubit data patches connected to a quantum bus using a square arrangement for performing lattice surgery. This is a time-efficient layout as all Pauli operators---$X$, $Z$, and $Y$ (the combination of both)---are directly accessible by the quantum bus. In an exact square arrangement, the $Q$ logical qubits requested in the algorithm would require a space of $2Q + \sqrt{8Q} + 1$ tiles plus the tiles required for the buffer space. To save on space, we shorten the last column when $\sqrt{Q/2}$ is not an integer. Different core layouts, such as the compact block layout of Ref.~\cite{litinski2018game} and the layout of Ref.~\cite{beverland2022surface} for edge-disjoint path compilation, can be incorporated into our models by adjusting the core processor size accordingly. However, a significant drawback of these layouts is their inability to provide access to the $Y$ operator of data qubits within a single logical cycle, which commonly appear in circuits after applying the transpilation procedure described in Refs~\cite{silva2024multi} and~\cite{litinski2018game}. This limitation requires converting logical gates with $Y$ operators into gates with only $X$ and $Z$ operations and commuting the additional Clifford gates out to avoid additional overhead in terms of time.

The buffer in \cref{fig:core_area} is designed to efficiently implement the post-corrected $\pi/8$ rotation protocol described in Ref.~\cite{litinski2018game} (see \Cref{app:pi8_protocols} for more details). Within this buffer, magic states produced by the MSF are sent to magic state growth patches where they are enlarged to match the code distance used in the core processor. Next, the correction state is created in the correction patch following the post-corrected protocol. The magic state used in the correction state creation is then moved to a magic state storage patch using patch deformation, while the correction qubit is moved to a correction storage patch, freeing their spaces for the next operation. Once stored, the magic state is ready for consumption by a $\pi/8$ rotation. To apply the rotation, the magic state is entangled with the data qubits in the memory fabric required for the operation using lattice surgery. After that, the $X$-basis single-qubit measurement of the post-corrected protocol is performed in the patch that was holding the magic state. At the earliest available time, the correction state entangled to it is measured in the $X/Z$-basis, determining whether a Pauli correction is required to define whether the rotation should have been a $+\pi/8$ or $-\pi/8$ rotation. This Pauli correction is not executed physically as it can be commuted to the end of the circuit, updating the axis of subsequent rotations based on whether they anticommute.

The time required to classically process these measurement results and determine subsequent operations is referred as the \textit{reaction time}. The reaction time determines whether $\pi/8$ rotation computation is $T$-count-limited or reaction-time-limited. In $T$-count-limited computation, each $\pi/8$ rotation is executed sequentially (or in parallel if subsequent rotations commute following a compilation procedure~\cite{silva2024multi}). In reaction-time-limited computation, quantum teleportation can be used to parallelize non-commutable $\pi/8$ rotations when the reaction time is faster than the time needed to complete a logical cycle
(see \Cref{app:pi8_protocols} for further details).

The number of correction storage patches in the buffer depends on the reaction time, as a patch remains occupied until this classical processing has completed. The buffer is designed to allow the simultaneous preparation of magic and correction states without introducing delays. If the MSF produces magic states sufficiently fast to keep the buffer continuously active, a magic state will always be available for consumption at the start of every logical cycle. 

Additional buffers can be placed around the memory fabric to enable the parallel preparation of multiple magic states, supporting parallel $\pi/8$ rotations. If the MSF production rate is slower than one magic state per logical cycle, the buffer size can be adjusted accordingly to minimize space requirements.

\subsection{Magic State Factory} \label{sec:magic_state_factory}

\begin{figure}[h]
    \centering
    \includegraphics[width=\linewidth]{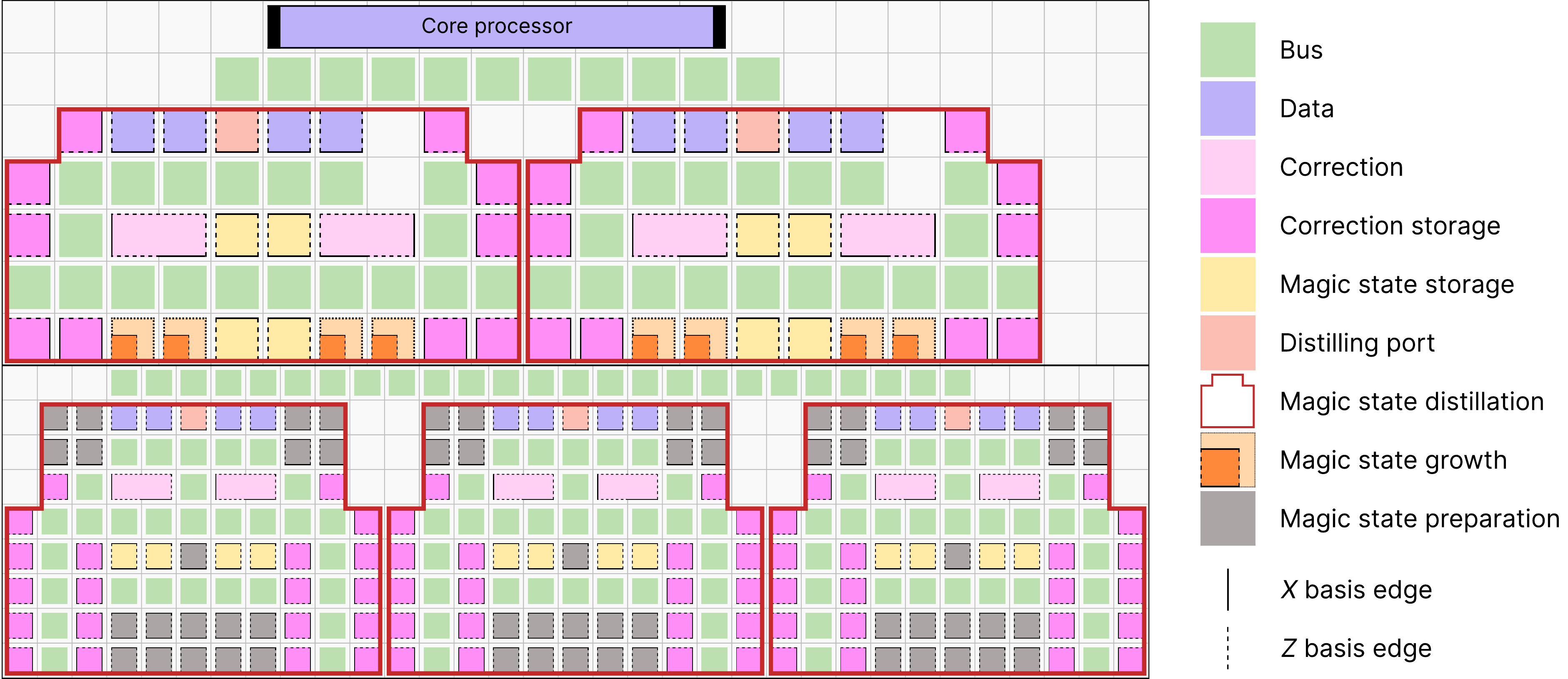}
    \caption{{\bf Architecture of the MSF and its interface with the core processor buffer.} Magic states are prepared using dedicated preparation units following a preparation protocol. Once preparation succeeds, magic states wait until a storage patch is available and then are moved to there using patch deformation. When in a storage patch, a similar post-correction protocol as is performed in the buffer follows, where a correction qubit is created, entangled with the magic state, and stored. The magic state is then used for the $\pi/8$ rotations performed between this magic state and data and distilling port qubits within the distillation unit. The higher-fidelity magic states are prepared in the distilling ports of these units. Once ready, they proceed to growth units dedicated to a next-level distillation unit. The process repeats until the highest level sends a magic state to the core processor. In this example, two distillation levels are represented for 15:1 distillation protocols, composed of three and two distillation units each from lowest to highest. While the lower-level unit design allows distillation of higher-fidelity magic states every 13 logical cycles if enough preparation units are available, topological limitations make the higher-level unit distill magic states only every 15 logical cycles instead due to the extra time required to load magic states into the data qubits before a distillation cycle begins.}
    \label{fig:magic_state_factory}
\end{figure}

The MSF is a module of the fault-tolerant architecture dedicated to the production of high-fidelity magic states required to implement the $\pi/8$ rotation gates. Magic states are initially prepared using physical operations in a \textit{magic state preparation unit}. As these are highly error-prone states that most likely have fidelities below those required for fault tolerance, rounds of magic state distillation are performed using \textit{magic state distillation units}, where the magic states at one particular level are consumed to create higher-fidelity magic states. Once ready, the magic states distilled are transferred to a distillation unit in the next level where they are expanded to a larger code distance, if required, in \textit{magic state growth units}. When magic states reach the required fidelity, the distillation process has been completed and the magic states created at the highest distillation level are sent to the core processor to be consumed when performing the $\pi/8$ rotation gates. \Cref{fig:magic_state_factory} shows an example of the architecture described.

The entire process of magic state distillation is a logical protocol that enables the consumption of lower-fidelity magic states to produce a smaller number of higher-fidelity magic states with some probability of success. There are several such protocols with varying trade-offs between them, such as length, the amount of increase in fidelity, the number of required input magic states, and the number of output magic states~\citep{bravyi2012magic,meier2012magicstatedistillationfourqubitcode,fowler2019lowoverheadquantumcomputation}. However, most known protocols are created using specific error correction or detection protocols in order to reduce the probability of specific types of errors.

Since magic state distillation protocols are performed on logical qubits, each protocol must be performed at a specific code distance. Furthermore, since operations on the logical qubits are faulty, there will be a term in the output fidelity of the magic states dependent on this code distance. In general, this term can be decreased by increasing the code distance at which the magic state distillation operates, but there is generally a trade-off between decreased Clifford errors and increasing numbers of physical qubits operating in a given magic state distillation unit. While it is difficult to theoretically determine the additional error these faulty operations cause, numerical studies can be used to determine how the error rates in the Clifford operations affect the outputs~\citep{beverland2022assessing}.

As a good example that is used as a testbed throughout this paper, the 15:1 distillation protocol~\citep{bravyi2012magic} uses 15 lower-fidelity magic states and produces a single higher-fidelity magic state. An analysis of this distillation protocol shows that if the input error rates of the magic states are $e_{\text{in}}$ and the logical error rates of the individual logical gates are $e_{\text{mem}}$, then the output error is bounded by $O(e_{\text{in}}^3 + e_{\text{mem}})$ from the distillation protocol. From this, one can determine the output fidelity of a single magic state distillation unit given the input fidelities and code distances. Additionally, it is possible to approximate the probability of a successful distillation for this protocol using numerical simulations~\citep{beverland2022assessing}. \Cref{fig:magic_state_factory} shows an architecture for the 15:1 protocol. The first-level distillation units are customized to run in 13 logical operations, accounting for 11 gates, one cycle for correction, and one cycle for emptying the distillation port, at a given code distance. Higher-level distillation units require two additional logical steps to load the four data (or stabilizer) qubits before starting a new distillation cycle due to topological constraints. This architecture requires at least 28 surface code tiles per distillation unit, even without accounting for the correction storage and magic state preparation space, which is much greater than the five logical qubits required for the protocol itself.

The very first level of the multi-level MSF consumes logical magic states prepared from physical non-Clifford states that are fault-tolerantly grown to the logical state of a full surface code patch. These protocols generally involve the creation of a physically relevant state (such as a physical magic state), and then the injection of this magic state into a larger quantum error-correcting code~\citep{Li_2012magic_state_injection,gidney2023cleaner}. The result is a magic state of fidelity $e_{\text{prep}}$ at a specified code distance. Different protocols require different physical operations, and have different time scales, but most require $\sim$1--2 logical cycles, where a logical cycle requires $d$ rounds of QEC at the final code distance $d$. The physical execution of the protocols on a 2D layout causes additional error to be accumulated which can also be numerically estimated. At the end of the process, we obtain the error rate $e_{\text{in}}$ for the initial level of magic state distillation. These preparation protocols generally have a probability of failure, and these probabilities can be numerically estimated.

Although distillation levels can be encoded using the same code distances, it is more space efficient to use different encodings for the logical qubits at different levels, usually in increasing order. Therefore, a \textit{growth protocol} must be used between the MSF levels to increase the code distance of the magic states to match the distance of the next distillation level or that of the core processor. These growth protocols operate similarly to the preparation protocols, in which a state encoded in a smaller state is surrounded by additional qubits in specific states, followed by QEC rounds in the grown patch. These growth protocols also have a small probability of failure that can be numerically estimated.

Magic state distillation is a probabilistic procedure, and thus each distillation unit has some chance of failing to output a state. To reduce the probability of no magic state existing when one is needed in our proposed architecture, we can set additional distillation units to ensure the system operates in a steady state, ensuring the average production is above consumption. This can be complemented with the inclusion of a buffer register between levels where magic states can remain idle while waiting to be consumed without blocking the start of the next distillation cycle in the unit where it was produced. These zones essentially need only to be a single logical tile where the magic state can sit, but having access to the tiles to both load and unload the magic states means they should generally be lined up.

\section{The Assembly Process} \label{sec:exec_time}

Designing the fault-tolerant described quantum architecture  requires solving a bi-objective optimization problem that balances minimizing the allocated space (i.e., the physical qubits required) and time (i.e., the runtime for circuit execution) under a given error budget. The decisions to be made are related to sizing the components of the architecture---specifically, the core processor and the MSF---such that both objectives are optimized while ensuring fault tolerance.

The assembler receives an overall error budget for the execution of a given quantum circuit composed of multi-qubit $\pi/8$ rotation gates. This error budget is then distributed between errors that arise in the execution of quantum operations in the core processor, $E_{\text{core}}$, and in the production of magic states in the MSF, $E_{\text{msf}}$. Therefore,
\begin{equation}\label{eq:E_prev}
    E_{\text{core}} + E_{\text{msf}} \leq E.
\end{equation}
The error models detailed in \Cref{app:error-model} provide a breakdown of this error budget as
\begin{equation}\label{eq:E_post_main_text}
    e_{\text{msf}}T + e_{\text{mem}, L+1} \left(KQ - \sum\limits_{i = 1}^{T^+} q_i\right) + \sum\limits_{i = 1}^{T^+} e_{\text{surg}, i} \leq E.
\end{equation}
where $e_{msf}$ represents the error rate of each magic state consumed in the core processor, $e_{\text{mem}, L+1}$ accounts for logical memory errors in the core processor, and $e_{\text{surg}, i}$ denotes the error contribution from each lattice surgery operations in the core processor for a compilation with $K$ logical time steps (makespan) for all $T$ non-Clifford and $T^+ - T$ Clifford operations in the circuit. The number of logical data qubits $q_i$ among all $Q$ data qubits involved in the logical quantum program is subtracted from the total to avoid double-counting the data qubits that are actively involved in a lattice surgery (see also \Cref{app:variables_and_parameters} for a
summary of all variables and parameters used in our error models). In this
section, we show how the sizing of the MSF and all code distances involved are
decided such that the accumulated errors from all quantum operations in the
core processor and in the MSF are below $E$ while balancing space and time
costs.

\subsection{Space and Time Costs} \label{sec:space-time-costs}

The space and time costs with respect to the architecture depend on the sizes of the core processor and the MSF, as well as the protocol used for the computation. Ignoring the warm-up time (i.e., the time needed to start filling the core buffer with high-quality magic states and prepare the correction qubits in the buffer), which is negligible for utility-scale quantum circuits, we can calculate the runtime $R$ of the quantum circuit as
\begin{equation}
    R = K \max{\{Wd_{L+1}, \gamma\}}.
\end{equation}
In a $T$-count-limited computation, the makespan $K = T$ logical time steps, and each step will take the longer between a logical cycle time in the core processor, that is, $d_{L+1}$ parity checks, each taking a time $W$ determined by hardware characterization for the QEC scheme considered, and the reaction time $\gamma$. Idling logical cycles also require performing $d_{L+1}$ parity checks to protect the important data. In a reaction-time-limited computation, which is only optimal when $\gamma < Wd_{L+1}$ (see~\Cref{sec:reaction_time_sensitivity}), $K = \delta \lceil T/(n-1) \rceil$ for an architecture with $n$ memory fabrics and each of the $\lceil T/(n-1) \rceil$ $BP$-$P$-$BM$ rounds taking $\delta$ logical time steps. For a given hardware device, $R$ is minimized by reducing the product of the code distance of the core processor $d_{L+1}$ and the makespan $K$, which is determined based on the production rate of magic states in the MSF and the demand profile for magic state consumption in the core.

Transpilation procedures can reduce circuit depth---and consequently the makespan---by commuting and eliminating Clifford gates from a quantum algorithm~\cite{silva2024multi}. This optimization smooths fluctuations in magic state demand, enabling a more consistent consumption rate. Compilation methods, such as those described in Refs.~\cite{silva2024multi} and \cite{beverland2022surface}, identify sets of $\pi/8$ rotation gates that can be performed in parallel. These approaches typically assume an uninterrupted supply of magic states for all gates scheduled within the same logical cycle.

Let a quantum algorithm composed of $T$ non-Clifford gates be represented by $T = \sum_{i=1}^{T_{\text{depth}}} t_i$, where $T_{\text{depth}}$ denotes the minimum number of logical cycles required to execute the algorithm (the shortest possible makespan, $K \leq T_{\text{depth}}$), and $t_i$ is the maximum number of $\pi/8$ rotations executable in time step $i$, as determined by the compilation. By assuming a transpiled algorithm with only non-Clifford gates and a steady magic state supply system and ignoring demand fluctuations since transpiled circuits are generally less parallelizable, the core processor’s maximum consumption rate can be defined as
\begin{equation}\label{eq:core_consumption_rate}
    C_{L+1} = \dfrac{Wd_{L+1}T_{\text{depth}}}{T},
\end{equation}
since logical cycles would be expected to consume, on average, $T/T_{\text{depth}}$ magic states every logical cycle that takes a time $Wd_{L+1}$. For the reaction-time-limited computation, the consumption rate of each core processor is slowed down by a factor of the time required for the Bell state preparations and basis measurements. On average, assuming $T = T_{depth}$, $n-1$ magic states are consumed across all $n$ core processors every $\delta$ time steps.

The production rate of magic states in a factory is determined by the production rate of all distillation units in its uppermost level, that is, $D_{L}$. A time-optimal architecture requires
\begin{equation}\label{eq:time_optimal_condition}
    D_{L} \geq C_{L+1}.
\end{equation}
Different scenarios are observed:
\begin{itemize}
    \item If $D_{L} > C_{L+1}$, then the MSF is oversized, which increases the space cost but with no reduction in the time cost, except when the magic state demand is too volatile, since they can be stocked during periods of low demand to avoid a shortage during periods of high demand;
    \item If $D_{L} = C_{L+1}$, then the MSF is space efficient to minimize the time cost; and
    \item If $D_{L} < C_{L+1}$, then the MSF is undersized, so the circuit takes longer to execute. The added idling time to the makespan $K$ inflates the idling volume $V_{\text{idle}}$, as defined in \cref{eq:idling_volume}. Thus, although the number of distillation units in the MSF decreases, the core processor size may increase due to there being a lower error budget availability for the errors accumulated during lattice surgery, as shown in \cref{eq:E_core}.
\end{itemize}
To achieve the production rate $D_{L}$ required for a time-optimal solution, we must ensure that production and consumption rates are balanced over all distillation levels. The production rate of any distillation level $l \in \{1, \ldots, L\}$ composed of $u_l$ parallel distillation units for this level is bounded by
\begin{equation}\label{eq:production_rate}
    D_l \leq \dfrac{u_l N P}{W d_l O},
\end{equation}
where $N$ is the number of magic states produced in a distillation round, $P$ is the probability of success for magic state distillation, $Wd_l$ is the execution time of each logical operation in the unit, and $O$ is the number of logical cycles required for the distillation round. The parameters $N$ and $O$ are given based on the distillation protocol, the success probability is a function \mbox{$P = f_{\text{acc}}(e_{\text{in},l}, e_{\text{mem},l})$,} and $W$ is based on the hardware characterization. Equality in \cref{eq:production_rate} is obtained when a steady execution of logical operations is possible due to a sufficient supply of magic states. Given that the consumption rate of magic states by all units in this level is
\begin{equation}\label{eq:consumption_rate}
    C_l \leq \dfrac{u_l M}{W d_l O},
\end{equation}
a steady flow is possible when $D_{l-1} \geq C_{l}$, where $M$ is the number of input magic states for the distillation unit. For example in the 15:1 distillation protocol $M = 15$, $N = 1$, $O = 13$ (i.e., 11 logical cycles for the $T$ gates in the protocol, one logical cycle for the corrections, and one cycle to empty the distilling port and position the lower-fidelity magic states for the next distillation cycle), and
\begin{equation}\label{eq:succ-prob-15-1}
P = 1 - 15e_{\text{in}} - 356e_{\text{cliff}}.
\end{equation}

To achieve a space-efficient steady flow, we need to ensure that
\begin{equation}\label{eq:optimal_msf_sizing}
    D_{l-1} = C_{l}, \quad \forall l \in \{1, \ldots, L+1\}.
\end{equation}
Combining the previous equations, we obtain the following relations between the number of distillation units and code distances between levels, which are given for the uppermost, intermediate, and lowermost levels, respectively, as
\begin{align}
    \dfrac{u_L N P}{d_L O} &= \dfrac{d_{L+1}T_{\text{depth}}}{T},\label{eq:time_optimal_L}\\
    \dfrac{u_{l-1} N P}{d_{l-1}} &= \dfrac{u_{l} M}{d_{l}}, \quad \forall l \in \{2,\ldots, L-1\},\label{eq:time_optimal_l}\\
    D_0 &= \dfrac{u_{1} M}{d_{1}}.\label{eq:time_optimal_0}
\end{align}
In the lowermost level, the magic state preparation rate $D_{0}$ is determined based on the preparation unit cycle time according to the magic state preparation protocol implemented.

Solving \crefrange{eq:time_optimal_L}{eq:time_optimal_0} considering the error budget in \cref{eq:E_post_main_text} leads to an architecture that approximates the space required to minimize time costs. Space costs are measured by the number of physical qubits required to encode all logical qubits in each area of the layout. Let the total space cost $S$ be represented as the sum of the physical qubits required at each area $l$, i.e.,
\begin{equation}
    S = \sum\limits_{l=1}^{L+1} S_l.
\end{equation}
The number of logical qubits in the core processor is determined as described in \Cref{sec:core_area} for the memory fabric, that is, $2Q + \sqrt{8Q} + 29$ logical qubits when $Q$ is a perfect square and slightly fewer logical qubits otherwise. We consider buffers with a size of 18 logical qubits (represented in \cref{fig:core_area}), which guarantees a steady flow of magic states to the memory fabric at a rate of one magic state per logical cycle in the time-optimal case. The correction storage required is determined by converting the reaction time $\gamma$ into logical cycles and accounting for accessibility, approximated as one bus patch for every two additional correction storage patches. Therefore, given that each logical qubit in the surface code requires $2d^2 - 1$ physical qubits, the core processor requires approximately 
\begin{equation}
    S_{L+1} = (2Q + \sqrt{8Q} + 47 + 1.5\lceil \gamma/(d_{L+1}W) \rceil) (2d_{L+1}^2 - 1)
\end{equation}
physical qubits.

The total space cost for the MSF is given by the space required for each distillation level. In the 15:1 distillation unit layout used in \cref{fig:magic_state_factory}:
\begin{itemize}
    \item each first-level distillation unit requires 29 logical qubits, plus additional qubits for magic state preparation and correction storage. Three preparation areas are used: the core requires a magic state prepared per time step, while the two upper areas require one every $O$ logical cycles. Using a steady-state approximation, the required number of preparation units per area is $n_{\text{prep}} = s \times (c_{\text{prep}}+p_{\text{prep}}) / p_{\text{prep}}$, where $p_{\text{prep}}$ is the preparation unit success probability, $c_{\text{prep}}$ is the preparation cycle time, and $s$ the required production rate;
    \item each higher-level distillation unit requires 28 logical qubits, plus correction storage; and
    \item each distillation level $l$ still requires at least $10(u_l-1) + 1$ qubits ($11u_l + 1$ for the first level) to connect the distillation units to the upper-level growth units.
\end{itemize}
Based on these considerations, the total space cost for the first level in the MSF is
\begin{equation}
    S_1 = u_1 (40 + 15/13(c_{\text{prep}}+p_{\text{prep}}) / p_{\text{prep}} + 1.5\lceil \gamma/(d_{L+1}W)\rceil) - 10,
\end{equation}
and for each of the higher levels $l = \{2,\ldots,L\}$, it is
\begin{equation}
    S_l = u_l (38 + 1.5\lceil \gamma/(d_{L+1}W) \rceil) - 9.
\end{equation}
While solving \crefrange{eq:time_optimal_L}{eq:time_optimal_0} leads to a time-optimal solution, if a distillation unit can meet the requirements to produce higher-fidelity magic states from lower-fidelity ones, only one distillation unit per level is necessary for an MSF to be able to produce magic states at the error rates meeting the condition in Eq.~(\ref{eq:E_post_main_text}). As previously mentioned, an MSF with $u_l = 1, \forall l \in \{1,\ldots, L\}$, may significantly reduce the size of the MSF $\sum_{l =1}^{L} S_l$ at the cost of increasing $S_{L+1}$ through an increase in $d_{L+1}$. This trade-off is analyzed in detail in \Cref{sec:experiments}.

\subsection{Our Assembly Method} \label{sec:solution}

Although the problem presented in \Cref{app:error-model} can be solved analytically, the complexity of representing the nested function for the accumulated MSF errors in \cref{eq:E_post_main_text} and the relations between the decision variables in \crefrange{eq:time_optimal_L}{eq:time_optimal_0} demand the development of a fast heuristic algorithm to solve the problem. This algorithm generates layouts for the considered architecture under both time-optimal and time-suboptimal scenarios.

Since the objective is to balance space and time costs, the slowdown factor $\beta$ we introduce acts as a target value for the runtime slowdown by choosing an undersized MSF. The number of logical cycles required to run the quantum algorithm is approximated by $K = \beta T$. A time-optimal solution corresponds to the expression $\beta = T_{\text{depth}}/T$, while the value $\beta = 1$ leads to a space-optimal solution that leads to approximately no excess idling time for the serial scheduling case. For $\beta > 1$, the MSF must be sized accordingly to estimate the resulting idling time. While the slowdown may not match the value of $\beta$ precisely due to the discrete nature of distillation units, we adjust $K$ accordingly to match the actual slowdown resulting from the MSF size chosen to approximate $\beta$.

We now detail the decision-making process for solving the optimization problem described. Note that the parameters $\alpha$ and $\beta$ are present in \cref{eq:E_core_approximation} as factors of $V_{\text{act}}$ and $V_{\text{idle}}$, respectively. As such, they are set prior to solving the problem to approximate the error rates in the core processor.

\subsubsection{Set the Core Area's Code Distance}

The first decision to be made is to determine the code distance $d_{L+1}$ for the core processor, assuming $E_{\text{msf}} = 0$ (i.e., there are no magic state factory errors). The minimum code distance that satisfies \cref{eq:E_post_main_text} is found by solving the following problem:
\begin{align}
    \min \quad & d_{L+1}\\
    \text{s.t.}\quad & (V_{\textrm{idle}} + V_{\text{act}})e_{\text{mem},L+1} \leq E,\\
    &d_{L+1} \in \{2k+1 | k \in \mathbb{Z}_+\},
\end{align}
where $V_{\textrm{idle}}$, $V_{\text{act}}$, and $e_{\text{mem},L+1}$ are represented by the Eqs.~(\ref{eq:idling_volume}), (\ref{eq:active_volume}), and (\ref{eq:e_mem}), respectively. Since all the parameters except $d_{L+1}$ are known a priori, this problem is easily solved by gradually increasing $d_{L+1}$ within its feasible space until a feasible solution for the problem is found.

\subsubsection{Set the First-Level's Code Distance}

Once $d_{L+1}$ has been determined, we calculate $E_{\text{core}}$ using \cref{eq:E_core_approximation} and, following \cref{eq:E_prev,eq:E_msf}, we set the magic state error budget according to
\begin{equation}
    e_{\text{msf}} \leq \dfrac{E - E_{\text{core}}}{T}.
\end{equation}
If this error threshold can be achieved without distillation, that is, $e_{\text{prep}}$ is below the right-hand side threshold above, the problem has been solved since no distillation is required. Otherwise, we proceed by defining a first distillation level and selecting the code distance $d_1$ that minimizes the magic state error rate:
\begin{align}
    \min \quad & e_{\text{out},1}\label{eq:model2_of}\\
    \text{s.t.}\quad & P > 0,\label{eq:model2_acc_prob}\\
    & d_1 \in \{2k+1 | k \in \mathbb{Z}_+\}.
\end{align}
Here, the objective function (\ref{eq:model2_of}) is a function for the magic state output error rate as in \cref{eq:dist,eq:dist-15-1}, and the constraint (\ref{eq:model2_acc_prob}), determined for example by \cref{eq:succ-prob-15-1}, ensures that the success probability is positive in order to validate the choice. While this problem can be solved using gradient-based methods, the limited solution space affords a quick way of verifying optimality by iterating through code distances while verifying feasibility until the turning point is reached in both functions $e_{\text{out},1}$ and $P$.

\subsubsection{Determine the Number of Distillation Levels}

Minimizing the number of distillation levels $L$ generally results in more space-efficient solutions by avoiding the overhead associated with additional levels, even those with small code distances. The magic states produced by the MSF must achieve an error rate below the threshold defined in \cref{eq:E_msf}, that is, $e_{\text{out},L} \leq e_{\text{msf}}$. Considering the accumulated errors described in \Cref{sec:msf-error-model}, the error rate of the output magic states after a distillation round is bounded by $35e_{\text{in}, l}^3$ for the 15:1 distillation protocol under the assumption that $d_l \rightarrow \infty$, since $e_{\text{cliff}, l} \rightarrow 0$ as derived from \cref{eq:cliff-error-model,eq:dist-15-1}. Consequently, to minimize $L$, it suffices to incrementally add distillation levels with $d_l = \infty$ until the magic state output error rate satisfies the required threshold.

\subsubsection{Update the Code Distances for All Levels}

Once the number of distillation levels has been set, we update the code distances $d_l$ for all levels, including the first one. This is because we are not looking to improve the error rate at each level as much as possible, but instead only to improve it so as to meet the error budget $E$. Beginning with the first level, we search for the smallest value of $d$ that leads to the error budget being met when updating the error rates in \cref{eq:e_msf} throughout the distillation process. This iterative process continues until all levels meet the error budget.

\subsubsection{Determine the Number of Distillation Units}

Finally, we determine the number of distillation units, $u_l$, at each level. We first calculate the consumption rate in the core $C_{L+1}$ using \cref{eq:core_consumption_rate} and the maximum production rate $D_{L}$ of a single distillation unit $u_L = 1$ at the level $L$ based on \cref{eq:production_rate}. Then, the number of distillation units is set to
\begin{equation}
    u_L = \left\lceil \dfrac{D_{L}}{\beta C_{L+1}} \right\rceil.
\end{equation}

This process is repeated for each level using \cref{eq:consumption_rate} for the consumption rate of each level in the MSF with $\beta$ adjusted based on the production rate of the selected number of units. Thus, when the chosen number of distillation units $u_l$ leads to a production rate above the target $\beta$, $u_{l-1}$ takes into consideration this gap and selects only enough units to approximate the target slowdown rate.

\subsubsection{Local Search}

To search for further improvements to the solution, a local search is conducted by incrementally increasing the code distance at each level, one level at a time, rebalancing production and consumption rates, and recalculating space costs. If an improvement is found, the solution is updated, and the process continues. The search stops when no further local improvements are possible. This process enables our optimization framework to always find optimal solutions for cases where $L \leq 2$, which correspond to the most common architectures in the experiments we conducted. For $L \geq 3$, the local search can be replaced with a greedy search that checks all possible combinations of code distances, ensuring optimality, if desired, but requiring additional classical computation.

\section{Numerically Studied Space--Time Trade-offs} \label{sec:experiments}

We conducted a set of experiments to analyze optimal solutions for the studied problem and to perform a sensitivity analysis on the parameters that were input, including the number of logical qubits ($Q$), the circuit depth ($T$), and hardware noise parameters. For simplicity we assumed $e_{\text{cliff}} = e_{\text{mem}}$ and that the quantum memory protocol had $r= d$ rounds everywhere, therefore, the hardware noise parameters were merely the quantum memory error suppression rate ($\Lambda_{\text{mem}}$), and the preparation error rate ($e_{\text{prep}}$). The experiments did not require much computational power; they should be easily reproducible even on a personal computer. We also assumed the 15:1 distillation protocol with the layout of \cref{fig:magic_state_factory} were used for all $\forall l \in \{1,\ldots,L\}$; therefore, $M = 15$, $N = 1$, $O = 13$, $P = 1 - 15e_{\text{in}} - 356e_{\text{mem}}$, and $e_{\text{out}} = 35e_{\text{in}}^3 + 7.1e_{\text{mem}}$.

As part of the default settings, we considered the error budget $E = 1\%$. We use the ``Target'' transmon qubit array parameters presented in Ref.~\cite{mohseni2024build} as an example (see their Table 1 for $T_1$ and $T_2$ times, and gate, preparation, measurement, and reset times, and their respective error rates). We perform circuit-level noise model simulations of rotated surface codes with these specifications using 1QBit's TopQAD~\citep{1qbit2024topqad} and obtain $\Lambda_{\text{mem}} = 9.3$ and $\mu_{\text{mem}} = 1.9 \times 10^{-2}$ and a magic state preparation error rate of $e_{\text{prep}} = 4.73 \times 10^{-5}$. The parity-check circuits for this hardware take $W = 350$ nanoseconds and we set the reaction time to $\gamma = 10\ \mu s$. All experiments were conducted with $\alpha = 0.1$, as preliminary results showed that variations in $\alpha$ only had negligible impacts on the results.

\subsection{Space and Time Cost Estimates}

\begin{figure}[t]
    \centering
    \includegraphics[width=0.55\linewidth, trim=0 0 0 50, clip]{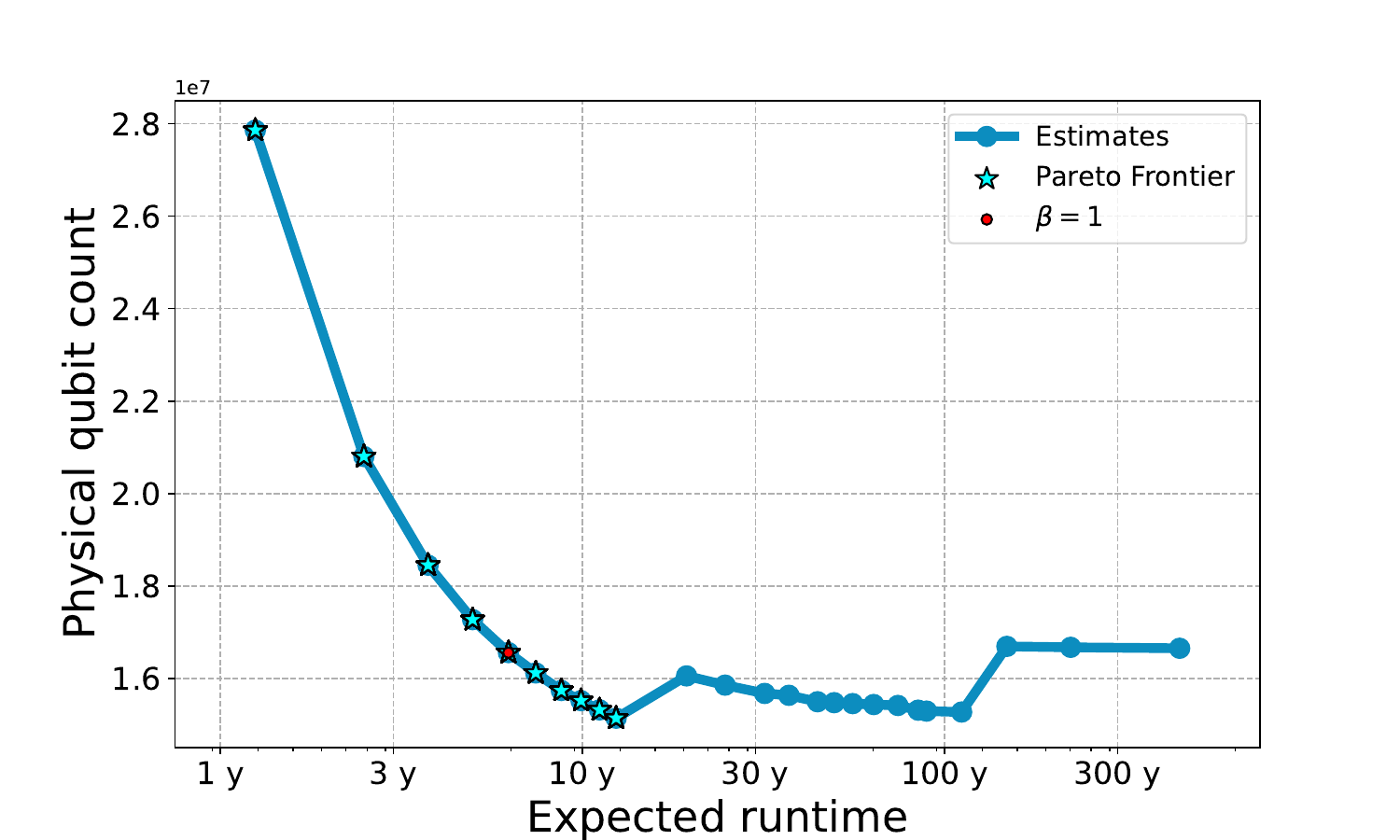}
    \caption{{\bf Space and time cost estimates for varying magic state factory sizes.} Estimates for the FeMoco-76 circuit with $T = 1.4 \times 10^{13}$ and $Q = 1972$m and quantum computers with $\Lambda_{\text{mem}} = 9.3$. The Pareto frontier points are highlighted, indicating the optimal solutions for the assembly problem. The point where $\beta = 1$ represents the serial scheduling case. The points to the left of $\beta = 1$ assume different degrees of circuit parallelization, while those to the right reflect increasing $\beta$ values until all distillation levels have only one unit. The two sharp increases in the physical qubit counts are due to increases in code distances in the core processor needed to meet the error budget at extended idling times.}
    \label{fig:pareto_front}
\end{figure}

We provide space and time cost estimates for electronic-structure quantum computations for the 76-orbital FeMoco molecule (FeMoco-76) with a precision in energy estimation of 0.1 mHa generated using Double Factorization Qubitization, as presented in Ref.~\cite{mohseni2024build}. The resulting quantum circuit is estimated to require the execution of $1.4 \times 10^{13}$ $T$ gates using 1972 logical data qubits. This circuit serves as a case study to demonstrate the capabilities of our methods. It is important to note that electronic-structure quantum computation is an active field of research, and circuits with reduced logical resource requirements may be achievable using alternative algorithms or by relaxing chemical accuracy constraints~\cite{low2025fast}.

\Cref{fig:pareto_front} presents the space and time cost estimates for the FeMoco-76 circuit. Each point in the figure represents a solution to the assembly problem for a predefined value of $\beta$, ranging from $\beta = 0.2$ to the case where each distillation level has only one unit. The Pareto frontier is highlighted, showing the optimal trade-offs between space and time costs. The solutions on this frontier range from 1.2 years and 27.2 million physical qubits (for $\beta = 0.2$) to 12.4 years and 15.1 million physical qubits (for $\beta \approx 2$).

In the best-case time scenario, assuming a highly parallelizable circuit (for $T_{\text{depth}} = 0.2T$), a $5\times$ speedup could be achieved with 66\% more qubits compared to serial execution (for $\beta = 1$). The trend in the plot demonstrates that physical qubit counts increase rapidly as runtime is further reduced, indicating how costly it is to implement the extra resources needed in an MSF to produce magic states sufficiently quickly to allow performing operations in the core processor in parallel whenever possible.

Looking in the other direction, the space-optimal solution reduces the number of physical qubits by 8.2\% compared to serial scheduling, and it slows down execution by a factor of $2\times$. Both solutions require two distillation levels (i.e., $L = 2$), but, for $\beta = 1$, the number of distillation units is $u \in \{72,14\}$, whereas for $\beta = 2$, it is reduced to $u \in \{36,7\}$. This represents a significant reduction in MSF size---from around 2.7 million to 1.3 million physical qubits---given the code distances $d \in \{15, 37\}$. The two spikes observed for the physical qubit estimates in the plot of \cref{fig:pareto_front} reflect increases of $d_{L+1}$ from 41 to 43 and then to 45.

\subsection{Circuit Parameter Sensitivity}

\begin{figure}[t]
\centering
\begin{subfigure}{0.49\textwidth}
    \centering
    \includegraphics[width=\linewidth, trim=0 0 0 50, clip]{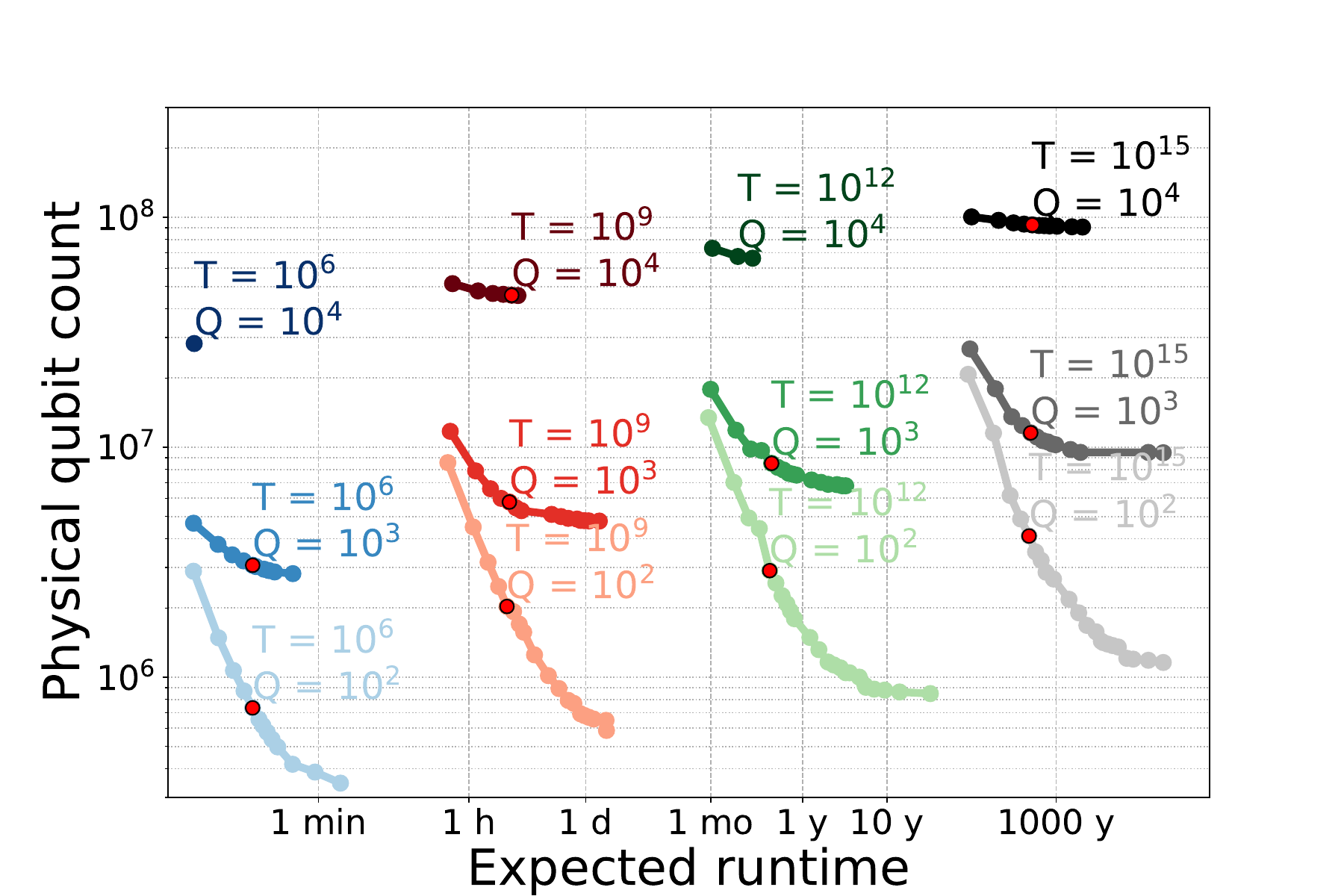}
    \caption{Space--time trade-offs with varying circuit parameters.}
    \label{fig:results_a}
\end{subfigure}
\hfill
\begin{subfigure}{0.49\textwidth}
    \centering
    \includegraphics[width=\linewidth, trim=0 0 0 50, clip]{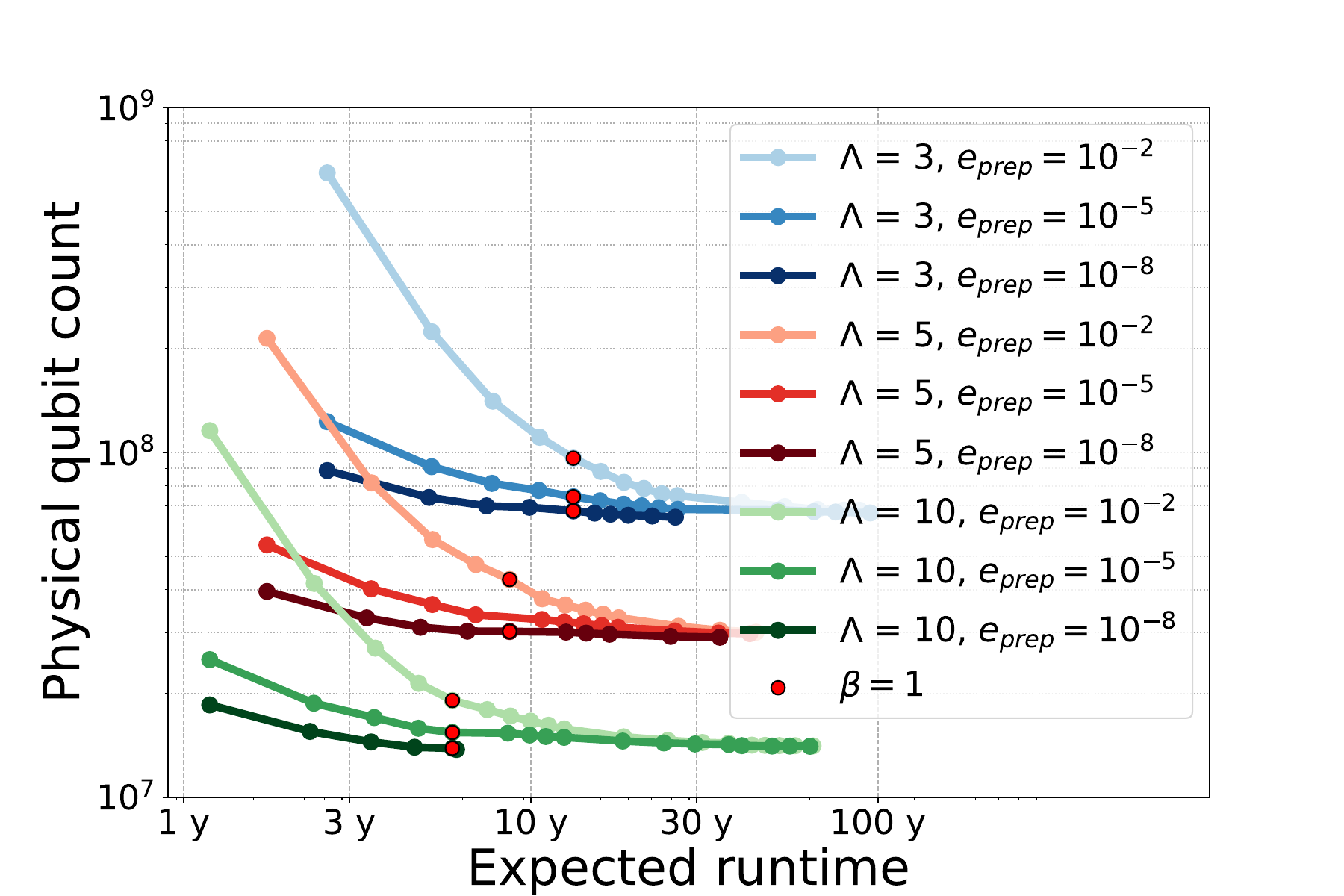}
    \caption{Space--time trade-offs with varying hardware parameters.}
    \label{fig:results_b}
\end{subfigure}

\caption{(a) Pareto frontier estimates are shown for combinations of $T$-count $T = 10^6$ to $10^{15}$ and logical data qubits $Q = 10^2$ to $10^4$ in a quantum circuit, based on an FTQC model with a quantum memory error prefactor of $\mu_{\text{mem}} = 1.9 \times 10^{-2}$, an error suppression rate of $\Lambda_{\text{mem}} = 9.3$, and a magic state preparation error rate of $e_{\text{prep}} = 4.73 \times 10^{-5}$. While runtime is heavily influenced by the number of non-Clifford gates in the circuit and the MSF size, these estimates demonstrate that quantum computers require between $10^5$ and $10^8$ physical qubits across all scenarios evaluated. (b) Pareto frontier estimates for combinations of $\Lambda_{\text{mem}} = 3$ to $10$ and a magic state preparation error rate of $e_{\text{prep}} = 10^{-2}$ to $10^{-8}$ for the FeMoco-76 circuit with $T = 1.4 \times 10^{13}$ and $Q = 1972$. The estimates show that improving $\Lambda_{\text{mem}}$ among the scenarios tested can reduce the physical qubit count by up to one order of magnitude with a smaller impact on the expected runtime due to the reduced code distances. Meanwhile, improving $e_{\text{prep}}$ has a small effect on reducing physical qubit count because of the relatively large size of the core processor compared to the MSF, and no effect on the expected runtime.}
\end{figure}

To illustrate how different circuits' characteristics affect resource estimates, \cref{fig:results_a} presents space and time cost estimates for varying circuit sizes in the ranges $T \in \{10^6, 10^9, 10^{12}, 10^{15}\}$ and \mbox{$Q \in \{10^2, 10^3, 10^4\}$.} These magnitudes are applicable to a broad spectrum of quantum applications. Only the Pareto frontier points are shown for each circuit size.

For every 1000-fold increase in the value of $T$, the runtime for the case where $\beta = 1$ increases by a factor of approximately 1000. One important observation is that few space-optimal solutions exist below the serial scheduling case when $Q = 10^4$. This is because, in large core processors, increasing the core processor's code distance makes the overall space costs worse than does saving space by having a smaller MSF. However, in smaller core processors (for $Q = 10^2$), there are opportunities to reduce space costs by downsizing the MSF. In such cases, space can be saved by up to an order of magnitude, while larger core sizes barely affect space costs.

Interestingly, when analyzing circuits of different sizes, we observe that the space requirements for running relatively smaller or larger circuits are comparable. For example, for $Q = 10^2$, the space cost for $T = 10^6$ ranges between $10^5$ and $10^6$ physical qubits, while for $T = 10^{15}$, the space requirements increase by less than an order of magnitude. In the cases where serial scheduling is employed, physical space requirements increase from approximately $6 \times 10^5$ to $4 \times 10^6$ qubits.

\subsection{Hardware Noise Parameter Sensitivity}

Next, we performed experiments to analyze the sensitivity of space and time estimates to changes in the hardware noise profile, particularly, by varying the error suppression factor $\Lambda_{\text{mem}}$ and magic state preparation error rate $e_{\text{prep}}$. \Cref{fig:results_b} shows the results for the FeMoco-76 circuit, considering the combinations of $\Lambda_{\text{mem}} \in \{3, 5, 10\}$ and preset values of the magic state preparation error rates $e_{\text{prep}} \in \{10^{-2}, 10^{-5}, 10^{-8}\}$. Only the Pareto frontier points are shown for each circuit size.

The figure shows that, under a scenario where the circuit has a parallelization potential leading to a $5\times$ speedup compared to the serial case, a circuit with $T = 1.4 \times 10^{13}$ $\pi/8$ rotations could be run in 1--3 years depending on the quality of the physical qubits and the prepared magic states. The bottleneck in terms of time is the increased code distance in the core processor due to lower-quality qubits. For example, the worst-case scenario ($\Lambda = 3, e_{\text{prep}} = 10^{-2}$) requires a 2.2 times higher encoding for the core processor than the best-case scenario ($\Lambda = 10, e_{\text{prep}} = 10^{-8}$), that is, $d_{\text{core}} = 87$ in the former and $d_{\text{core}} = 39$ in the latter, both for the $\beta = 0.2$ scenario. This difference also affects space requirements---\mbox{6.8 times} fewer physical qubits are needed in the best-case scenario as compared to the worst (13.9 million vs. 94.4 million).

\begin{figure}[t]
     \centering
     \begin{subfigure}[b]{0.49\columnwidth}
         \centering
         \includegraphics[trim=0 0 0 50, clip, width=1.0\linewidth]{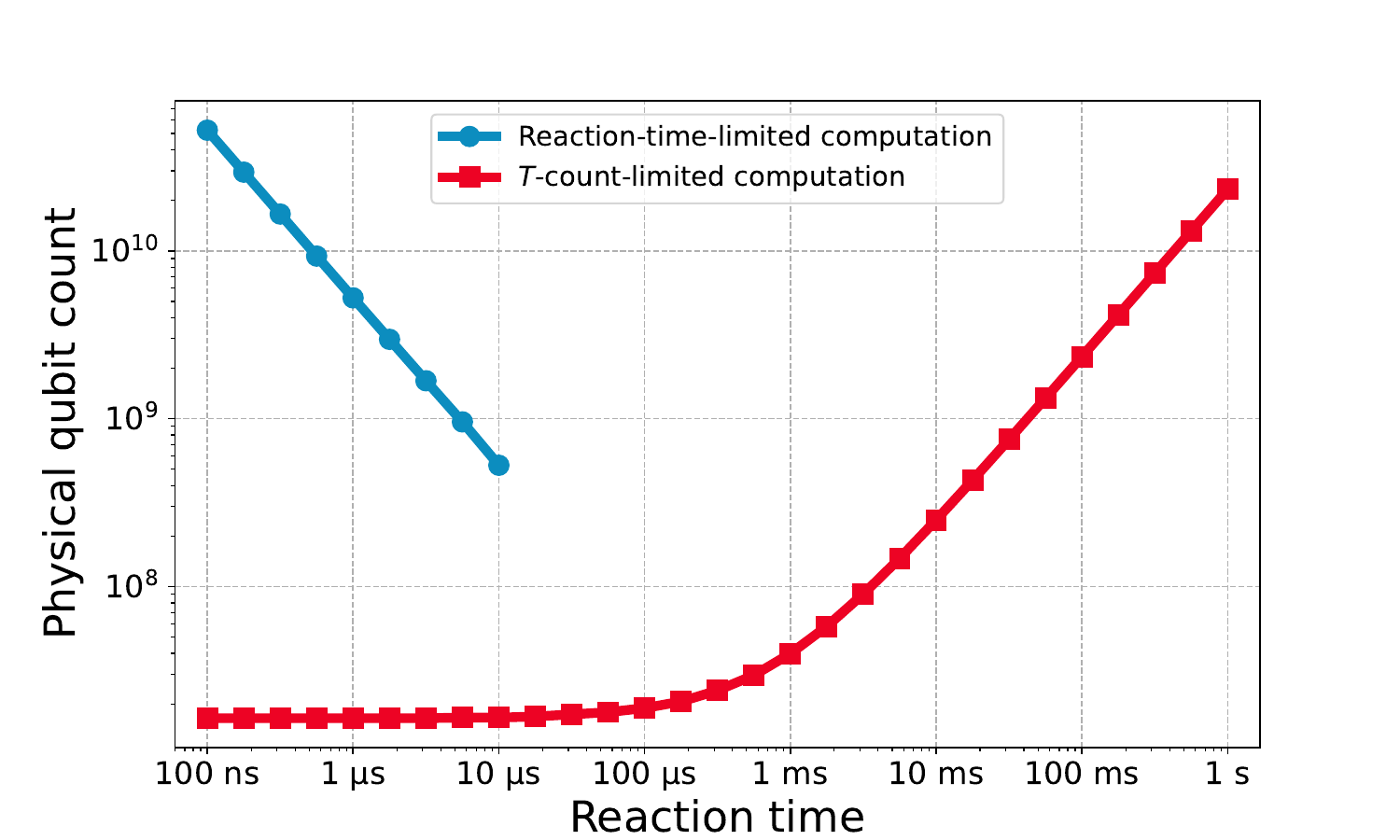}
         \caption{Space costs}
         \label{fig:rtl_a}
     \end{subfigure}
     \hfill
     \begin{subfigure}[b]{0.49\columnwidth}
         \centering
         \includegraphics[trim=0 0 0 50, clip, width=1.0\linewidth]{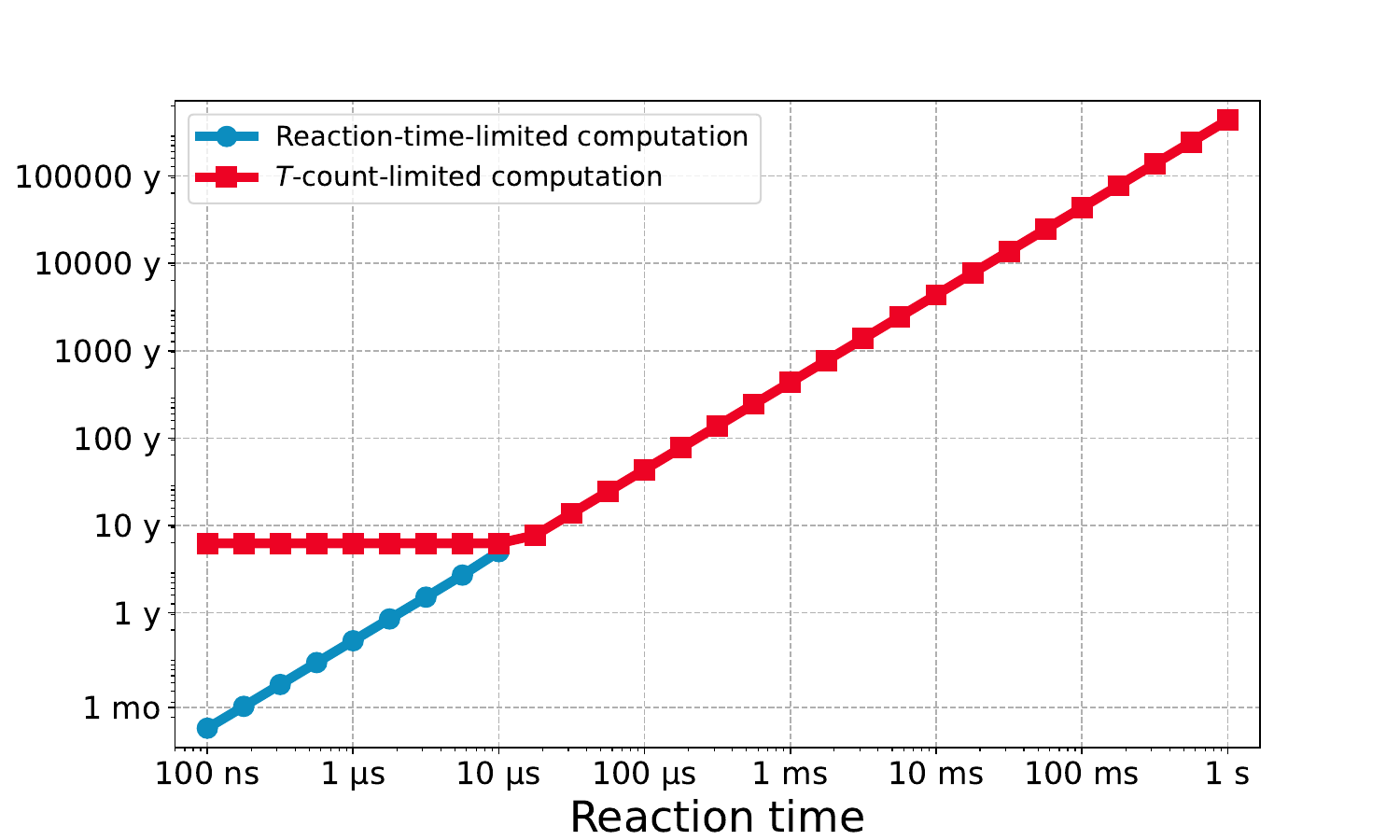}
         \caption{Time costs}
         \label{fig:rtl_b}
     \end{subfigure}
     \caption{{\bf Space and time cost estimates for varying reaction times.} Estimates for the FeMoco-76 circuit with $T = 1.4 \times 10^{13}$ and $Q = 1972$, and a quantum computer with $\Lambda_{\text{mem}} = 9.3$, for $\beta = 1$. Estimates are generated for reaction times $\gamma$ ranging from 100 ns to 1 s. The plots indicate that, for the hardware specification assumed, reaction-time-limited computation provides speedups when $\gamma \lesssim d_{L+1}W$, which, for the scenario tested with $d_{L+1} = 41$ and $W = 350$ ns, results in the threshold at $14.35 \mu$s. Beyond this threshold, runtime decreases linearly with decreasing reaction time at a linearly increasing space cost. For slower reaction times, $T$-count-limited computation is optimal. However, longer reaction times increase space costs due to the larger correction qubit storage requirements in the buffer and MSF.}
     \label{fig:rtl_sensitivity}
\end{figure}

\subsection{Reaction Time Sensitivity} \label{sec:reaction_time_sensitivity}

We conducted experiments to analyze how space and time estimates vary with reaction time, that is, the time required to classically process measurement outcomes and determine corrective operations in $\pi/8$ rotations. \Cref{fig:rtl_sensitivity} presents the space and time cost estimates obtained, distinguishing between estimates for the $T$-count-limited computation and the reaction-time-limited computation (as described in \Cref{app:pi8_protocols}). The threshold where reaction-time-limited computation can provide a speedup is approximately $\gamma = d_{L+1}W$. Estimates for the $T$-count-limited computation converges to the $\beta = 1$ estimates in \Cref{fig:pareto_front} as reaction time $\gamma \to 0$.

In practical quantum systems, large delays in performing quantum error correction can degrade the error suppression rate for quantum memory. At a certain point, slow communication between quantum and classical systems could push quantum error correction above threshold, i.e., $\Lambda_{\text{mem}} < 1$, rendering fault-tolerant computation infeasible. The effects of the reaction time on quantum error correction can be emulated numerically and incorporated into the noise models presented by modelling noise as a function of reaction time, that is, $f_{\text{mem}}(d,r,\gamma)$.

\section{Conclusion} \label{sec:conclusion}

In this paper, we have presented a method for designing a fault-tolerant quantum architecture that optimizes the trade-offs between the space and time costs needed to execute large-scale quantum circuits. Our approach integrates a core processor designed for efficient logical qubit operations with a multi-level magic state factory that supplies high-fidelity magic states, enabling an effective implementation of non-Clifford gates, which is required for universal quantum computation. We developed a fast heuristic algorithm to optimize the architecture under different FTQC protocols and circuit and hardware parameter settings, providing a flexible framework for estimating quantum resource requirements.

The main takeaway from our experiments is that quantum resource estimates can be quickly generated when a few key parameters are known, such as the circuit volume (in terms of $\pi/8$ rotation gates and logical qubits requested), the error suppression rates of the quantum error correction scheme, the reaction time, and the specific protocols used for magic state preparation, growth, and distillation. Although certain simplifying assumptions were made to facilitate model development, such as using the same distillation protocols across all distillation levels, the proposed framework is adaptable to other quantum architectures, including alternative MSF designs and other resource state factories.

We observed that, depending on the circuit and hardware characteristics, the quantum error correction scheme in the proposed architecture requires between 100,000 and 100 million physical qubits, scaling to higher numbers when reaction time becomes significantly faster than the logical cycle times to provide computational speedups. When comparing our approach to previous resource estimations in the literature, we found consistent physical resource estimates for utility-scale quantum algorithms with runtimes ranging from seconds to days. Future research may focus on discovering novel quantum algorithms, improving circuit synthesis and compilation, and drastically improving hardware quality to further reduce physical requirements.

Several avenues for future work have arisen from this study. First, improved FTQC protocols can change the resulting resource estimates, which may affect some of the conclusions we have drawn. Additionally, our concept architecture in this paper has a free-space layout, but constraints such as the size of individual QPUs and quantum interconnects between multiple QPUs for distributed quantum computing can be incorporated. Finally, our approach can be applied to design other resource factory modules beyond the $\ket{T}$ states, such as Toffoli states, QROMs, and resource states for specific rotation angles, further broadening the applicability of our approach to quantum architecture design.

\section*{Acknowledgement}
The authors thank our editor, Marko~Bucyk, for his careful review and editing of the manuscript.
The authors are grateful to Craig~Gidney, Christopher~Chamberland, John~Martinis,
\linebreak
Masoud~Mohseni, and Alan~Ho for useful discussions.
The authors acknowledge the financial support of Pacific Economic Development Canada (PacifiCan) under project number PC0008525.
G.~A.~D.~is grateful for the support of Mitacs.
P.~R.~acknowledges the financial support of Mike and Ophelia Lazaridis, Innovation, Science and Economic Development Canada (ISED), and the Perimeter Institute for Theoretical Physics. Research at the Perimeter Institute is supported in part by the Government of Canada through ISED and by the Province of Ontario through the Ministry of Colleges and Universities.
\bibliography{references}

\clearpage
\appendix

\section{Protocols for the implementation of fault-tolerant $\pi/8$ rotations} \label{app:pi8_protocols}

The implementation of fault-tolerant $\pi/8$ rotations requires executing a protocol to detect and correct errors when consuming a magic state. For that, we implemented the post-corrected $\pi/8$ rotation protocol described in Ref.~\cite{litinski2018game}.

In this protocol, a resource state, referred to as a \textit{correction state}, is prepared by initializing a qubit in the $\ket{0}$ state and performing a $Z \otimes X$ measurement between the magic state and the created correction state. The $\pi/8$ rotation is then applied by measuring $P \otimes Z$ between the data qubits involved in the operation and the magic state being consumed. Afterwards, the consumed magic state is measured in the $X$ basis, and the correction qubit is measured either in the $X$ or $Z$ basis, depending on the outcome of the $P \otimes Z$ measurement, to determine the corrective operation to be performed.

\begin{figure}[h]
    \centering
    \includegraphics[width=\linewidth]{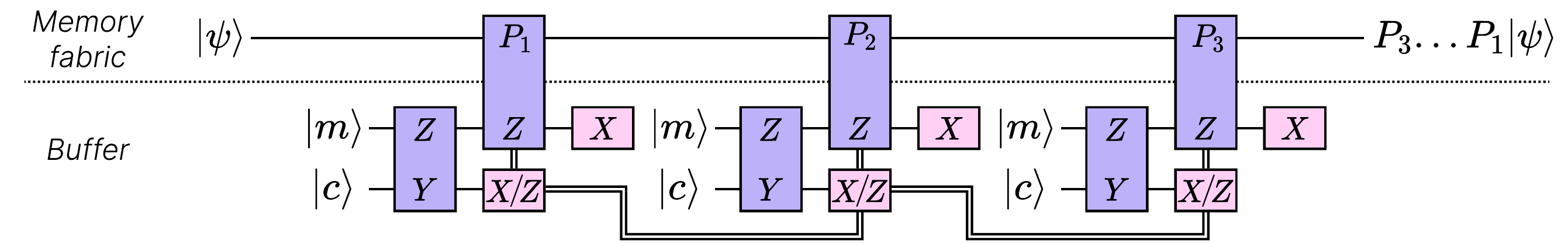}
    \caption{{\bf Protocol for a $T$-count-limited computation.} Each operation $P$ representing a $\pi/8$ rotation is performed in sequence using the post-corrected protocol. A magic state $\ket{m}$ created externally is used to create a correction state $\ket{c}$ with a $Z \otimes X$ measurement. Then, a $P \otimes Z$ measurement is made for the data qubits in a state $\ket{\psi}$ and the magic state $\ket{m}$. This measurement outcome determined by a classical computation defines the basis ($X/Z$) for the correction qubit $\ket{c}$ measurement. This classical computation can be performed in parallel to the next $\pi/8$ rotation measurements except for the next correction qubit measurement, since previous $\ket{c}$ measurements affect corrections. This means that the time needed to perform $T$ $\pi/8$ rotations is limited by the $T$ count given there is sufficient space to store correction qubits awaiting classical processing.}
    \label{fig:T_count_limited_computation}
\end{figure}

The post-corrected protocol allows the implementation of $\pi/8$ rotations in serial. The time for each $\pi/8$ rotation is determined by two factors: the time needed to process the $P \otimes Z$ measurement using lattice surgery and the reaction time required to determine the basis of the correction qubit measurement. \Cref{fig:T_count_limited_computation} illustrates this $T$-count-limited computation. In this protocol, the correction qubit measurement for one operation can be performed in parallel with the $P \otimes Z$ measurement of the subsequent operation. As a result, if the reaction time is shorter than the time required to measure $P \otimes Z$, the computation is not slowed down by classical processing. However, if the reaction time exceeds this threshold, it becomes the limiting factor, defining the overall runtime of the computation.

In the $T$-count-limited computation, the total runtime is dictated by the sequence of measurements required for each operation. When the reaction time is sufficiently fast, quantum teleportation can be used to parallelize non-commutable operations, accelerating computation~\cite{fowler2013time}.  Quantum teleportation enables multiple non-commutable operations to be performed simultaneously by teleporting qubit states to auxiliary qubits entangled as Bell pairs.

\begin{figure}
    \centering
    \includegraphics[width=0.9\linewidth]{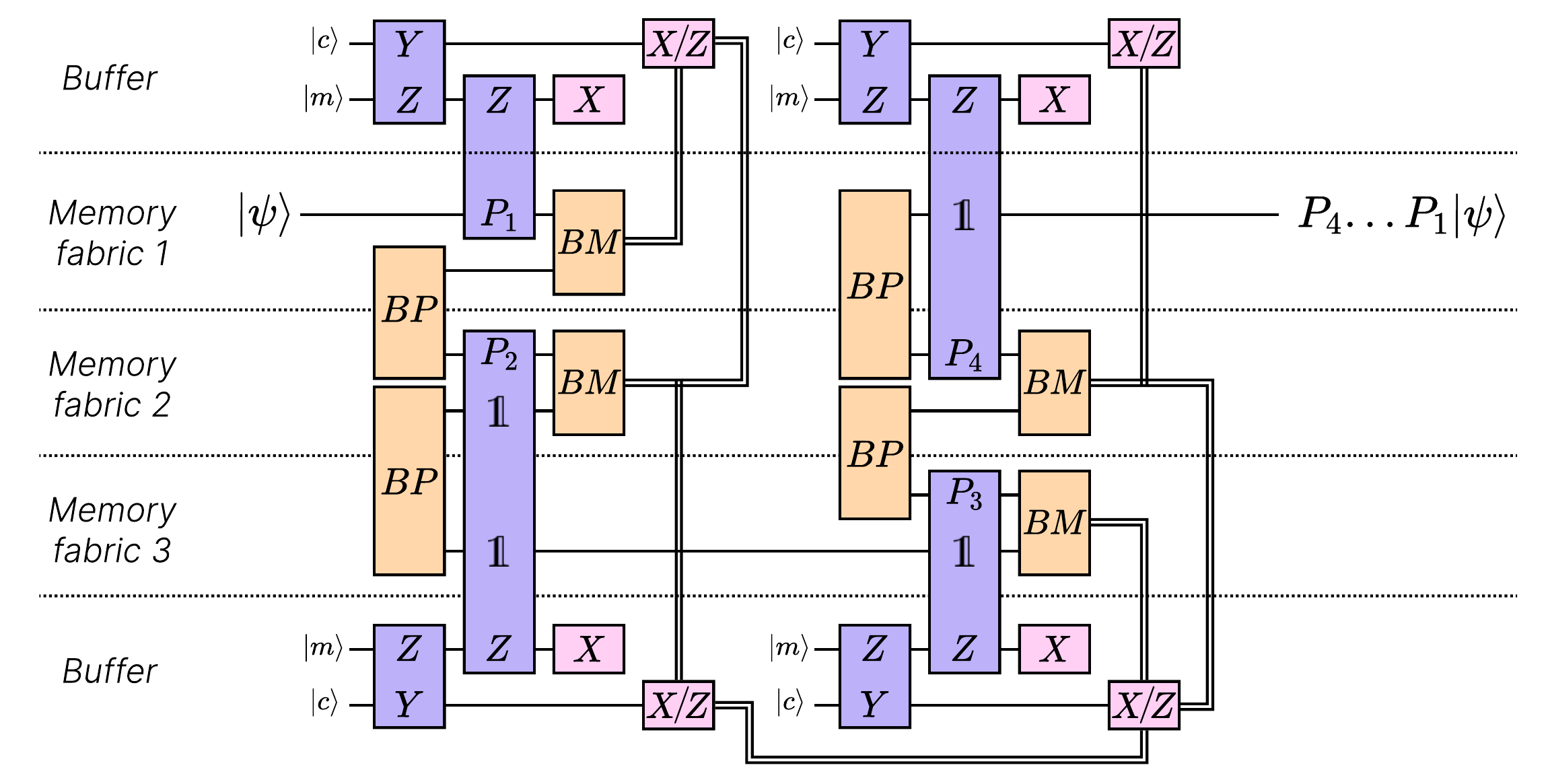}
    \caption{{\bf Protocol for space-efficient reaction-time-limited computation.} The computation is structured across multiple core processors, each containing memory fabrics with $2Q$ logical data qubits, with each half represented by a line. Bell state preparation ($BP$) entangles data qubits between adjacent memory fabrics, enabling teleportation-based parallelization of non-commutable operations ($P$). After applying $P$, Bell basis measurements ($BM$) are performed to teleport the transformed states to the next memory fabric. Assuming $P$ are $\pi/8$ rotations, the buffer manages magic state $\ket{m}$ and correction $\ket{c}$ qubits used for post-correction of the $\pi/8$ rotations. The outcomes of $BM$ measurements and previous $\ket{c}$ measurements determine the basis ($X/Z$) for the next $\ket{c}$ measurement. Correction qubits from a previous $BP$-$P$-$BM$ round can be measured in parallel with the $BP$, $P$ and $BM$ operations of the next round. The protocol ensures efficient space usage by reusing discarded qubits for subsequent $BP$-$P$-$BM$ rounds. In this example, three memory fabrics are used to perform the four non-commutable operations in two $BP$-$P$-$BM$ rounds.}
    \label{fig:reaction_time_limited_computation}
\end{figure}

The protocol for implementing this computation is shown in \cref{fig:reaction_time_limited_computation}. It starts by constructing $n$ adjacent core processors, each containing memory fabrics with $2Q$ logical data qubits. The first memory fabric holds the initial state $\ket{\psi}$ in half of its data qubits. The other half is entangled with the adjacent memory fabric via Bell state preparation (operation $BP$), which allows the teleportation of the transformed states $P\ket{\psi}$ after applying the operations $P$. This $BP$ process is performed across all core processors.

Once Bell states have been prepared, each memory fabric except the last one executes the $P_i$ non-commutable operations using the data qubits entangled with the memory fabric above (operation $P$). Then, Bell basis measurements (operation $BM$) are performed on the transformed data qubits and their entangled counterparts, teleporting the quantum state to the next memory fabric and discarding the data qubits in the memory fabrics where $BM$s have occurred. After all $BP$-$P$-$BM$ operations have been performed, the last memory fabric will contain the computed result $P_{n-1} \ldots P_2 P_1 \ket{\psi}$.

The success of teleportation is determined by the $BM$ outcomes. For $P$ taken as a set of $\pi/8$ rotations, Pauli errors may arise from the teleportation operations, requiring a Pauli correction. Using the post-corrected $\pi/8$ rotation protocol, the needed correction is determined by measuring the correction qubit entangled to the magic state consumed by the $\pi/8$ rotation that produced it, which is stored in the buffer correction storage space in our architecture. The basis measured in the correction qubit ($X$ or $Z$) is determined by the outcomes of its related $BM$ and the outcomes of the previous correction qubit measurements. The time required to finish all operations parallelized is then the time needed to perform the $BP$-$P$-$BM$ operations ($\delta$) and the time required for the classical processing (reaction time, $\gamma$) of these measurements serially. Therefore, for a computation involving $T$ $\pi/8$ rotations, this reaction-time-limited computation can only provide a speedup if $\delta + \gamma T$ is faster than executing all $T$ $\pi/8$ rotations serially, that is, in the $T$-count-limited computation.

As shown in \cref{fig:reaction_time_limited_computation}, the reaction-time-limited computation can be performed in rounds where the computation performed in the buffer correction storage can be run in parallel to the $BP$-$P$-$BM$ operations. This allows us to save on space, since qubits discarded after $BM$ operations can be reused by the operations in the next round. The runtime needed to perform a round is the greatest between $\delta$ and $\gamma (n-1)$. Therefore, in the time-optimal, space-efficient case, the maximum number of memory fabrics that can be used without slowing down the computation is $n_{max} = \delta / \gamma + 1$ and the minimum space required is the size of the $n$ core processors used, including their buffer correction storage, along with the infrastructure required to distill and distribute $n-1$ high-fidelity magic states using the MSF every $\delta$ logical time steps.

An efficient implementation of this reaction-time-limited computation is achieved using the fast block layout for the memory fabric. In this layout, $BP$ takes $\lceil\sqrt[4]{Q}\rceil + 1$ logical time steps, $P$ takes 1 logical time step, and $BM$ takes 2 logical time steps~\cite{litinski2018game}. Thus, $\delta = \lceil\sqrt[4]{Q}\rceil + 4$ logical time steps for a computation involving $Q$ data qubits. In \Cref{sec:reaction_time_sensitivity}, we analyze the space and time costs for different reaction times $\gamma$ in the time-optimal, space-efficient case. We note that using fewer than the maximum number of memory fabrics is also possible and also provides space--time trade-offs.

\section{Hardware Error Modelling} \label{app:error-model}

In this section, we detail how the errors of the core processor and the MSF are
modelled and predicted to provide a detailed breakdown of the constraint \cref
{eq:E_prev}. Our goal is to go beyond the simple quantum memory experiment
simulations typically used in resource estimators such as Azure's Quantum
Resource Estimator~\cite{beverland2022assessing} and Qualtran~\cite
{harrigan2024expressing}, and to incorporate the geometry of multi-qubit
lattice surgeries and the resulting time-like and space-like errors. Using
detailed circuit-level noise models, the errors produced by idling qubits, gate
infidelities, and SPAM (state preparation and measurement) errors are treated
independently \cite{mohseni2024build}. Additionally, the cumulative errors in
other FTQC procedures such as preparation, distillation, and growth stages of
magic state factories are taken into account.

\subsection{Core Processor Errors} \label{sec:core-error-model}

In the core processor, the quantum algorithm operates on $Q$ logical qubits to execute Clifford+$T$ gates over a makespan of $K$ logical cycles, where each cycle represents a layer of gates that are performed in parallel as defined in a lattice surgery compilation procedure, such as the one described in Ref.~\cite{silva2024multi}. The cycles are divided into \textit{active cycles}, where at least one gate is performed in the memory fabric, and \textit{idling cycles}, where the data qubits in the memory fabric must be preserved through error correction while waiting for resources (e.g., magic states).

During active cycles, for the execution of the operations scheduled, lattice surgeries connect data qubits to either a zero-state qubit in a Clifford operation or a magic state from the buffer in a non-Clifford operation. Lattice surgeries used to 
prepare magic states and correction qubits are also considered, although they are performed in earlier cycles---either active or idling ones. Each of these lattice surgeries requires performing a logical operation involving faulty logical qubits. The error probability for all lattice surgeries required to perform an operation with index $i \in \{1,\ldots, T^+\}$ is denoted as $e_{\text{surg}, i}$, where $T^+$ denotes the non-Clifford and Clifford operations in the circuit, and depends on the size and structure of the lattice surgeries performed~\citep{fowler2012surface}.

Logical data qubits containing valuable quantum information in the core that are not engaged in active operations must always be protected using rounds of parity checks either during active or idling cycles, which also contribute to the error accumulation. The \textit{idling volume} is defined as
\begin{equation}\label{eq:idling_volume}
    V_{\text{idle}} = KQ - \sum\limits_{i = 1}^{T^+} q_i,
\end{equation}
where $KQ$ represents all locations where a logical error can occur when protecting idling qubits, and $q_i$ is the number of logical data qubits involved in the logical quantum program, which is subtracted from the total to avoid double-counting the data qubits that are actively involved in a lattice surgery.

The probability of having at least one error when performing all logical operations for idling in the core processor is $1-(1-e_{\text{mem}, \text{core}})^{V_{\text{idle}}} \approx V_{\text{idle}}e_{\text{mem}, \text{core}}$, where $e_{\text{mem}, \text{core}}$ is the error rate of the logical qubits in the core processor. Consequently, the approximate total accumulated errors in the core processor from active and idling operations is
\begin{equation}\label{eq:E_core}
    E_{\text{core}} = V_{\text{idle}}e_{\text{mem}, \text{core}} + \sum\limits_{i = 1}^{T^+} e_{\text{surg}, i}.
\end{equation}
The above formula can be approximated as
\begin{equation}\label{eq:E_core-cliff}
    E_{\text{core}} = V_{\text{idle}}e_{\text{mem}, \text{core}} + T^+ e_{\text{cliff}, \text{core}},
\end{equation}
where $e_{\text{cliff}, \text{core}}$ represents the average error rates of all the multi-qubit lattice surgeries in the core. This approximation disregards the varying shapes of different lattice surgeries.

An alternative further simplification is to use only quantum memory error rates. To this end, we define the \textit{active volume} of a compiled program as
\begin{equation}\label{eq:active_volume}
    V_{\text{act}} = (2Q + \sqrt{8Q} + 26) \alpha T^+,
\end{equation}
where $(2Q + \sqrt{8Q} + 26)$ refers to the number of logical qubits in the core (i.e., the memory fabric and buffer), and $\alpha$ is the average size (i.e., the number of surface code patches) of the lattice surgeries in the program. This results in the approximation
\begin{equation}\label{eq:E_core_approximation}
    E_{\text{core}} \approx (V_{\textrm{idle}} + V_{\text{act}}) e_{\text{mem},\text{core}}.
\end{equation}

We assume that the assembler has access to the logical error rates of the QEC codes (at high distances). This information may be provided directly by performing experiments on the quantum hardware. Alternatively, a predictive (parameterized) model $f_{\text{mem}}$ can be regressed from numerical simulations at low distances using efficient stabilizer circuit simulators~\cite{aaronson2004improved, gidney2021stim}. To summarize, for a core processor of distance
$d_{\text{core}}$, we have
\begin{equation}\label{eq:core-error}
e_{\text{mem}, \text{core}}= f_{\text{mem}} (d_{\text{core}}).
\end{equation}

\begin{figure*}[t]

\begin{tabular}{cc}
  \begin{subfigure}{0.45\linewidth}
    \subcaption{Most-probable quantum memory errors}
    \includegraphics[width=\textwidth]{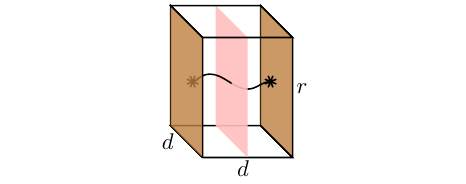}
    \label{fig:mem-bulk}
  \end{subfigure} \hspace{0cm} &

  \begin{subfigure}{0.45\linewidth}
    \subcaption{Most-probable $X$-type errors in surgery}
    \includegraphics[width=\textwidth]{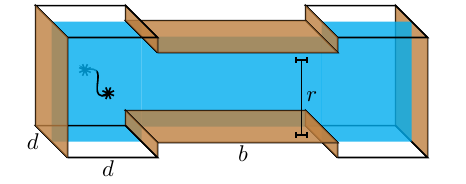}
    \label{fig:cliff-x}
  \end{subfigure} \hspace{0cm} \\

  \begin{subfigure}{0.45\linewidth}
    \subcaption{Most-probable $Z$-type errors in surgery}
    \includegraphics[width=\textwidth]{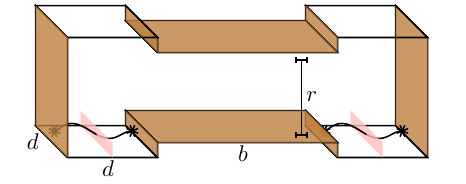}
    \label{fig:cliff-z}
  \end{subfigure} \hspace{0cm} &

  \begin{subfigure}{0.45\linewidth}
    \subcaption{Most-probable time-like errors in surgery}
    \includegraphics[width=\textwidth]{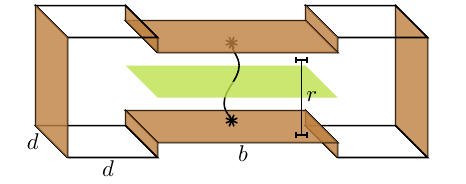}
    \label{fig:cliff-t}
  \end{subfigure}
\end{tabular}
\\[-2ex]
\caption{{\bf Shortest-length undetectable errors in FTQC protocols, which determine the leading factors of the predictive error models.} (a) For quantum memory experiments, both $X$- and $Z$-type errors land on boundaries of area $rd$ for $r$ rounds of stabilizer measurements. (b--d) For multi-qubit surgeries, no extra rounds of QEC are assumed at the split and the pre-merge phases. (b) Shortest-length error strings of the $X$ type have a cross section of area $O((2d+b)r)$. (c) Since no distinct pre-merge rounds are assumed, the area of the boundaries of the shortest-length $Z$-type errors grows as $O(d)$. \newline (d) The degeneracy of the shortest-length undetectable time-like errors scales with the area of the measured bus, $db$.}
\label{fig:area-law}
\end{figure*}

For a quantum memory experiment of distance $d$ involving $r$ QEC rounds, we use
\begin{equation}\label{eq:e_mem}
    f_{\text{mem}}(d, r) = \mu_{\text{mem}} dr \Lambda_{\text{mem}}^{-\frac{d+1}{2}}
\end{equation}
as our model, where the error prefactor $\mu_{\text{mem}}$ and the error suppression rate $\Lambda_{\text{mem}}$ are fitting parameters. The error suppression rate $\Lambda_{\text{mem}}$ represents the asymptotic factor of improvement in the logical fidelity of the QEC code when its distance is increased by 2. We expect $\Lambda_{\text{mem}} > 1$ for the surface code below threshold~\citep{fowler2012surface, kelly2015state, acharya2024quantum}. The exponential suppression with base $\Lambda$ represents the length $(d+1)/2$ of the most probable error strings, and the coefficient $dr$ accounts for the multiplicity of these error strings, governed by the area of the cross section of the code bulk with error strings of either $X$- or $Z$-type, as shown in \cref{fig:mem-bulk}. Typically, $r = O(d)$ to assure fault tolerance.

Similarly, the Clifford error rates $e_{\text{cliff}}$ can be approximated using stabilizer circuit simulations at low distances and extrapolated to higher distances. A model for the multi-qubit surgery errors can be extracted from a similar argument involving the shortest-length errors in the bulk as shown in \cref{fig:area-law}. Note that an area law does not apply to all error string types \cite{chamberland2022universal}. For example, in an $XX$ surgery between two logical qubits involving $r$ QEC rounds and a bus patch of length $b$, the error model is
\begin{equation} \label{eq:cliff-error-model}
    f_{\text{cliff}} (d, b, r) = \mu_X (2d+b)r \Lambda_X^{-(d+1)/2} + \mu_Z d \Lambda_Z^{-(d+1)/2} + \mu_T d b \Lambda_T^{-(r +1)/2},
\end{equation}
since the $X$-type errors have boundary surfaces of areas $r(2d+b)$ (see \cref{fig:cliff-x}), the $Z$-type errors grow with the length of the boundary edges shown in \cref{fig:cliff-z} and thus contribute to a linear factor of $d$ in the second term, and the time-like errors of the shortest distance land in a boundary of area $bd$ (see \cref{fig:cliff-t}).

\begin{figure}[t]
\includegraphics[width=0.4\textwidth]{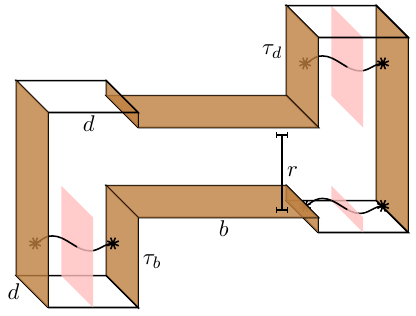}
\caption{{\bf Effect of buffer and decoder delays on surgery fidelities.} In core processor surgeries, a teleportation is implemented involving a magic state (represented by the code patch on the left) and data qubits (the code patch on the right). The magic state incurs a delay $\tau_b$ in the buffer, which in the space-efficient steady flow scenario is one logical cycle, on average (i.e., $d$ QEC rounds), before the surgery begins. The data qubit QEC rounds must continue until decoder decisions are available after a decoder delay time of $\tau_d$.}
\label{fig:buffer-decoder-delay}
\end{figure}

For teleportations in our core processor, the teleported state is a magic state that has resided in the buffer for some average expected buffer delay time, $\tau_b$ (which is 1 clock cycle for balanced production and consumption rates; see \cref{sec:space-time-costs} for further details). The targets of teleportation are logical data qubit patches in the core processor for which further QEC rounds are executed until the decoder outcome is available. We denote this delay by $\tau_d$. Inclusion of the buffer and decoder delays and assuming an average rate for all types of surgeries changes the above model to
\begin{equation} \label{eq:surg-error-model}
    f_{\text{surg}} (d, b, r) = \mu \big[d (2r + \tau_b + \tau_d + 1) + br \big] \Lambda^{-(d+1)/2} + \mu_T d b \Lambda_T^{-(r +1)/2},
\end{equation}
which still distinguishes time-like and space-like errors but ignores the type of surgery, for example, $XX$ merge (see \cref{fig:buffer-decoder-delay}).

\subsection{Magic State Factory Errors} \label{sec:msf-error-model}

The multi-level MSF described in \Cref{sec:magic_state_factory} produces the high-fidelity magic states required for the $\pi/8$ rotation gates. The probability of errors in the production of magic states affects the overall error rate of execution according to
\begin{equation}\label{eq:E_msf}
     E_{\text{msf}} = Te_{\text{msf}},
\end{equation}
where $T$ is the number of non-Clifford gates in the quantum algorithm and $e_{\text{msf}}$ is the error rate of the magic state consumed by each non-Clifford operation in the core.

In the MSF, an iterative process is followed that involves multiple levels of distillation to improve magic state fidelity. Each distillation level is linked with each other using \emph{growth zones}. Consider an MSF with $L$ levels: a $0$-th level referring to the magic state preparation area, the subset $\{1, \ldots, L\}$ referring to the consecutive distillation levels, and the $L+1$-st level referring to the core processor. The number of distillation levels $L \in \mathbb{Z}_{\geq 0}$ is to be determined such that the magic state error rate $e_{\text{msf}}$ meets the condition in \cref{eq:E_prev} while satisfying \cref{eq:E_msf}. The number of qubits used for building the MSF (the space cost) is optimized by choosing an increasing sequence $d_0, d_1, \ldots, d_{L+1}= d_{\text{core}}$ of code distances for each of the levels assuming a fault-tolerant code growth protocol is applied within the growth zones to promote code patches of distance $d_l$ to those of distance $d_{l+1}$.

The logical error rate of the magic states provided to the $l$-th MSF level is denoted by $e_{\text{in}, l}$ and the error rate of the resulting distilled magic state is $e_{\text{out}, l}$. Therefore, the contribution of the growth zones to the error rates of the magic states between levels is calculated via
\begin{equation}\label{eq:in}
    e_{\text{in},l} = g_{\text{grow}}(e_{\text{out},l-1}, e_{\text{grow}, l-1})
    = 1 - (1 - e_{\text{out},l-1})(1 - e_{\text{grow}, l-1}), \quad \forall l \in \{1,\ldots, L+1\},
\end{equation}
from simple probabilistic arguments where $e_{\text{grow}, l-1}$ is the error rate accumulated from the growth procedure performed after level $l-1$.

Similar to \cref{sec:core-error-model}, a predictive model (again, to be determined theoretically and regressed numerically, or, alternatively, experimentally) simulating the specific growth protocol is used to infer
\begin{equation}\label{eq:grow}
    e_{\text{grow}, l}= f_{\text{grow}}(d_l, d_{l+1})
\end{equation}
as a function of initial and final code distances, $d_l$ and $d_{l+1}$, in the growth zone. The output magic state error rates $e_{\text{out},l}$ are given by a function of the form
\begin{equation}\label{eq:dist}
e_{\text{out}, l}= g_{\text{dist}} (e_{\text{in}, l}, e_{\text{cliff}, l}),
\end{equation}
which, for example, for the 15:1 distillation protocol is
\begin{equation}\label{eq:dist-15-1}
e_{\text{out}, l}= 35e_{\text{in}, l}^3 + 7.1 e_{\text{cliff}, l}
\end{equation}
as per Refs.\cite{bravyi2005universal, litinski2018game} and \cite{beverland2022assessing}. The coefficient 35 is driven theoretically~\cite{bravyi2005universal}; however, the contribution of surgeries within the $l$-th distillation circuit is approximated as a numerically driven multiple (7.1 in Ref.~\cite{beverland2022surface}) of the Clifford error rate $e_{\text{cliff}, l}$ obtained from the model~\cref{eq:cliff-error-model}.

Overall, the magic state error rates of the entire MSF can be calculated recursively as follows:
\begin{align}
    e_{\text{out},0} &= e_{\text{in},1} = e_{\text{prep}}, \\
    e_{\text{out},l} &= g_{\text{dist}}(e_{\text{prep}}, e_{\text{cliff},l}), \quad \forall l \in \{1, \ldots, L\}, \\
    e_{\text{in},l+1} &= g_{\text{grow}}(e_{\text{out},l}, e_{\text{grow},l}), \quad \forall l \in \{1, \ldots, L-1\}, \\
    e_{\text{in},L+1} &= e_{\text{msf}} = g_{\text{grow}}(e_{\text{out},L}, e_{\text{grow},L}). \label{eq:e_msf}
\end{align}
Here, the preparation error rate $e_{\text{prep}}$ is estimated from numerical simulation of the fault-tolerant magic state preparation protocol in use at the zeroeth MSF level (i.e., at distance $d_0$). The numerical behaviour of these protocols as a function of $d$ in the error regimes in which we are interested can be quite distinct. For instance, for some protocols~\citep{singh2022high, gidney2023cleaner}, the error rate
\begin{equation} \label{eq:linear-prep}
    e_{\text{prep}} = f_{\text{prep}}(d_0)= \mu_{\text{prep}} - \Lambda_{\text{prep}}d_0,
\end{equation}
with a negative error suppression rate $\Lambda_{\text{prep}}$, which leads to the error rate increasing linearly with respect to distance. However, more-recent protocols~\citep{gidney2024magic} exhibit an exponential suppression in their error rates as far as numerical simulations in the regime of practically relevant distances and physical error rates are concerned:
\begin{equation}
    e_{\text{prep}} = \mu_{\text{prep}} \Lambda_{\text{prep}}^{-\frac{d+1}{2}}.
\end{equation}
In both cases, the starting factor $\mu_{\text{prep}}$ and the error suppression rate $\Lambda_{\text{prep}}$ are fitting parameters. Other protocols might have different behaviour altogether.

Combining Eqs.~(\ref{eq:idling_volume}), (\ref{eq:E_core}), and (\ref{eq:E_msf}), we can rewrite the error budget constraint \cref{eq:E_prev} as
\begin{equation}\label{eq:E_post}
    e_{\text{msf}}T + e_{\text{mem}, L+1} \left(KQ - \sum\limits_{i = 1}^{T^+} q_i\right) + \sum\limits_{i = 1}^{T^+} e_{\text{surg}, i} \leq E.
\end{equation}
Therefore, given a compiled quantum circuit with $T^+$ gates, among which $T$ are non-Clifford gates, involving $Q$ logical qubits, run (potentially with some parallelization) in $K$ logical cycles, involving $q_i$ qubits in the $i$-th operation, and the hardware error models described above, we can determine the number of distillation levels $L$, the code distances at each level, and the code distance in the core processor such that \cref{eq:E_post} is satisfied.

\section{Comparison with Azure's Resource Estimator}
\label{app:comparison}

We provide a comparison of our proposed architecture and resource estimates with the compilation scheme \textit{Parallel Synthesis Sequential Pauli Computation} (PSSPC) proposed in Beverland et~al.~\cite{beverland2022assessing}.

\textbf{Magic state factory.} The PSSPC also uses a multi-level MSF, where all distillation units within a level work synchronously and distillation between levels is performed serially using the same space. While this serial approach can save space compared to the parallel distillation between levels in our method, it comes with trade-offs. Serial distillation requires more units to mitigate the longer time required for distillation. Additionally, if distillation fails to produce the required number of magic states to supply the upper-level units, the process must restart or proceed with fewer than the required number of magic states to keep the production rate steady. This may lead to additional idling cycles in the core processor, increasing the computation time. Furthermore, the PSSPC's MSF design overlooks overheads related to magic state preparation, transfer, and growth. In our architecture, these processes increase the size of distillation units from the 20 logical qubits considered in the PSSPC's MSF design to an average of 34 logical qubits.

\textbf{Instruction set architecture.} Beverland et al. \cite{beverland2022assessing} considers circuits composed of $T$ gates, Clifford gates (logical measurements), Toffoli gates, and arbitrary-angle single-qubit rotation gates. While their implementation of $T$ gates is comparable to our $\pi/8$ rotations, the Toffoli gates and arbitrary-angle rotation gates require more-careful examination. Toffoli gates require four $T$ states to be consumed, but they can be scheduled to be executed in three logical cycles. In our optimization of physical resources, this Toffoli implementation can be represented by assuming a smaller slowdown factor $\beta$ to account for this three-cycle execution. Another optimistic assumption in Ref.~\cite{beverland2022assessing} is the parallel synthesis of parallelizable arbitrary-angle rotations. While this is good practice, they ignore both the synthesis qubits used for this process and the potential bus conflicts for the execution of these parallel operations. In our approach, such a parallelization potential is taken into consideration using a smaller slowdown factor $\beta$ or a shorter makespan $K$ without ignoring any physical resources, and the scheduling represented in Ref.~\cite{silva2024multi} can be used to assure no bus conflicts arise.

\textbf{Error model.} Similar to our paper, Beverland et~al.~\cite{beverland2022assessing} consider several physical qubit models. The most comparable model to our study is their ``($10^{-3}$, ns)'' model, which assumes physical error rates of $p= 10^{-3}$ with a surface code threshold of $p_{th}= 10^{-2}$, hence using a simplistic error suppression model $(p_{th}/ p)^{-(d+1)/2}$ when predicting logical error rates. We choose our $\Lambda = 9.3$ model for comparison with the results of Ref.~\cite{beverland2022assessing}. We note that the additional coefficient $dr$ in our memory error model \cref{eq:e_mem} leads to code distances with worse error rates. Therefore, the noise model in Ref.~\cite{beverland2022assessing} underestimates the required code distances. They also use a parity-check time of $W = 400$ ns, which is slightly longer than the one we use in this paper. Since reaction time is not discussed in Ref.~\cite{beverland2022assessing}, we present estimates for the case where $\gamma = d_{L+1}W$ to avoid the reaction-time-limited computation.

\textbf{Numerical comparison.} The logical and physical resource estimates for the three applications presented in Ref.~\cite{beverland2022assessing} (quantum dynamics, quantum chemistry, and factoring) are summarized in Tables~\ref{tab:logical_benchmark} and~\ref{tab:physical_benchmark}, respectively. Comparisons are provided for the $\beta = 1$ case. A comparison is also provided for the quantum dynamics application where the MSF is reduced to emulate a $\beta = 10$ case as assumed in Ref.~\cite{beverland2022assessing}.

\begin{table}[h]
\caption{Logical resources required for the quantum algorithms for real applications presented in Ref.~\cite{beverland2022assessing}.}
\centering
\resizebox{\textwidth}{!}{%
\begin{tabular}{@{}lcccc@{}}
\toprule
\textbf{Application} & \textbf{\# Logical qubits (\(Q\))} & \textbf{\# Magic states (\(T\))} & \textbf{Makespan (\(K\))} & \textbf{Error budget (\(E\))} \\
\midrule
Quantum dynamics & 100 & $2.4 \times 10^{6}$ & $1.5 \times 10^{5}$ & 0.001 \\
Quantum chemistry & 1318 & $5.4 \times 10^{11}$ & $4.1 \times 10^{11}$ & 0.01 \\
Factoring & 12,581 & $1.5 \times 10^{10}$ & $1.2 \times 10^{10}$ & 0.333 \\
\bottomrule
\end{tabular}
}
\label{tab:logical_benchmark}
\end{table}

\begin{table}[h]
\caption{Comparison of physical resource estimates for the quantum algorithms for real applications when using the PSSPC scheme from Ref.~\cite{beverland2022assessing} and our proposed architecture.}
\centering
\resizebox{\textwidth}{!}{%
\begin{tabular}{@{}lcc|cc@{}}
\toprule
 & \multicolumn{2}{c}{\textbf{PSSPC}} & \multicolumn{2}{c}{\textbf{Our architecture}} \\
\cmidrule(r){2-3} \cmidrule(l){4-5}
\textbf{Application} & \textbf{Physical qubits} & \textbf{Runtime} & \textbf{Physical qubits} & \textbf{Runtime} \\
\midrule
Quantum dynamics & $8.4 \times 10^6$ & $1.0$ s & $8.7 \times 10^5$ & $19.3$ s \\
Quantum dynamics (reduced MSF) & $9.2 \times 10^5$ & $13.5$ s & $3.5 \times 10^5$ & $251.3$ s \\
Quantum chemistry & $7.1 \times 10^6$ & $68.2$ d & $9.8 \times 10^6$ & $81.5$ d \\
Factoring & $3.9 \times 10^7$ & $1.7$ d & $7.1 \times 10^7$ & $2.2$ d \\
\bottomrule
\end{tabular}%
}
\label{tab:physical_benchmark}
\end{table}

\newpage
The physical resource estimates provided for our architecture consider a makespan $K = T$, representing a case where $\pi/8$ rotations are compiled with limited parallelization, such as when the circuit is transpiled to remove the Clifford gates, resulting in one magic state being consumed per logical cycle. The longer runtime of our estimates reflects the lack of parallelization of magic state consumption compared to the aggressive parallelization assumed in Ref.~\cite{beverland2022assessing}.

The differences in qubit count stem mainly from the different error models used as explained above. For example, in the factoring application, while Ref.~\cite{beverland2022assessing} predicts a code distance of $d_{L+1} = 27$ for the core processor, we assemble our core processor with a code distance of $d_{L+1} = 37$. This alone represents a memory fabric that is 87\% larger, leading to the higher estimates observed in the applications where the core processor size is dominant over the MSF size. When the MSF size is relatively larger, such as in the quantum dynamics application, although the benchmarking study omits MSF overheads resulting from magic state preparation, transfer, and growth, the reduced number of distillation units resulting from parallelization of distillation across levels in our MSF significantly offsets this difference, effectively reducing the MSF size. This analysis illustrates the several nuances and trade-offs that must be carefully considered when designing quantum architectures and performing resource estimations for FTQC.

\newgeometry{top=1in, bottom=-0.5in, left=0.65in, right=0.65in}
\section{Overview of Variables and Parameters}\label{app:variables_and_parameters}

\begin{xtabular}{cl}
    \toprule
    \textbf{Variable/Parameter} & \textbf{Description} \\ 
    \midrule

    \multicolumn{2}{l}{\textbf{Quantum Circuit}} \\ \midrule
    $T^+$ & Number of Pauli rotations (Clifford and non-Clifford) in the circuit \\
    $T$ & Number of $\pi/8$ rotations in the circuit \\
    $T_{\text{depth}}$ & Circuit depth \\
    $t_i$ & Number of $\pi/8$ rotations at each depth $i \in \{1,\ldots, T_{\text{depth}}\}$ \\
    $Q$ & Number of logical qubits in the circuit \\
    $q_i$ & Number of logical qubits requested by each $\pi/8$ rotation $i \in \{1,\ldots,T\}$ \\ \midrule

    \multicolumn{2}{l}{\textbf{Compilation}} \\ \midrule
    $V_{\text{idle}}$ & Idling volume \\
    $V_{\text{act}}$ & Active volume \\
    $C_{l}$ & Magic state consumption rate at level $l$ of the MSF \\
    $C_{L+1}$ & Magic state consumption rate at the core processor \\
    $D_{l}$ & Magic state production rate at level $l$ of the MSF \\
    $D_{0}$ & Magic state production rate at the preparation area \\ \midrule

    \multicolumn{2}{l}{\textbf{FTQC Emulation}} \\ \midrule
    $\gamma$ & Reaction time \\
    $W$ & Parity-check time \\
    $\mu_{\text{mem}}, \Lambda_{\text{mem}}$ & Quantum memory error prefactor and suppression rate\\
    $\mu_{\text{prep}}, \Lambda_{\text{prep}}$ & Magic state preparation error prefactor and suppression rate\\ \midrule

    \multicolumn{2}{l}{\textbf{Magic State Distillation}} \\ \midrule
    $M$ & Number of magic states required per distillation cycle \\
    $N$ & Number of magic states distilled per cycle \\
    $O$ & Number of logical cycles in a distillation cycle \\
    $P$ & Success probability of the protocol \\ \midrule

    \multicolumn{2}{l}{\textbf{Error Rates}} \\ \midrule
    $E$ & Error budget \\
    $E_{\text{msf}}$ & Magic state factory errors \\
    $E_{\text{core}}$ & Core processor errors \\
    $e_{\text{surg}}$ & Lattice surgery error rate\\
    $e_{\text{cliff}}$ & Average logical Clifford operation error rate\\
    $e_{\text{mem}, l}$ & Logical quantum memory error rate at level $l$ (MSF + core)\\
    $e_{\text{msf}}$ & Magic state error rate output from the MSF\\
    $e_{\text{in}, l}, e_{\text{out},l}$ & Magic state error rates input/output at level $l$ (MSF + core)\\
    $e_{\text{grow}, l}$ & Magic state error rate after growth at level $l$ (MSF)\\
    $e_{\text{prep}}$ & Magic state error rate after preparation \\ \midrule

    \multicolumn{2}{l}{\textbf{Costs}} \\ \midrule
    $R$ & Total expected runtime (time cost) \\
    $S$ & Total number of physical qubits (space cost) \\ \midrule

    \multicolumn{2}{l}{\textbf{Decision Variables}} \\ \midrule
    $\alpha$ & Average lattice surgery size relative to logical qubits in the core processor\\
    $\beta$ & Target slowdown factor\\
    $L$ & Number of distillation levels \\
    $K$ & Total logical cycles required to execute the algorithm (makespan) \\
    $d_l$ & Code distance for all logical qubits at level $l$ (MSF + core) \\
    $u_l$ & Number of distillation units used at level $l$ (MSF) \\

\end{xtabular}
\clearpage

\end{document}